\pdfoutput=1

\documentclass[11pt,twoside,a4paper,cmspaper,final,collab]{cms-tdr}
\def\svnVersion{389302}\def\cmsCernNoTag{CERN-EP/2016-152}\def\cmsCernDate{\today}\def\cmsMessage{Published in the Journal of High Energy Physics as \href{http://dx.doi.org/10.1007/JHEP02(2017)096}{\doi{10.1007/JHEP02(2017)096.}}}

\begin{document}\cmsNoteHeader{SMP-14-012}

\hyphenation{had-ron-i-za-tion}
\hyphenation{cal-or-i-me-ter}
\hyphenation{de-vices}
\RCS$Revision: 385817 $
\RCS$HeadURL: svn+ssh://svn.cern.ch/reps/tdr2/papers/SMP-14-012/trunk/SMP-14-012.tex $
\RCS$Id: SMP-14-012.tex 385817 2017-02-07 21:33:25Z alverson $

\providecommand{\PWpm}{\ensuremath{{\PW^\pm}}}%
\newcommand{\MN}{\ensuremath{\PGm\PGn}}%
\newcommand{\EN}{\ensuremath{\Pe\PGn}}%
\newcommand{\LN}{\ensuremath{\ell\PGn}}%
\providecommand{\MT}{\ensuremath{{M}_{\mathrm{T}}}}%
\newcommand{\PWmn}{\ensuremath{\PW \to \MN}}%
\newcommand{\PWpmn}{\ensuremath{\PWp \to \PGmp\PGn}}%
\newcommand{\PWmmn}{\ensuremath{\PWm \to \PGmm\PAGn}}%
\newcommand{\Wtn}{\ensuremath{\PW \to \PGt\PGn}}%
\newcommand{\Wen}{\ensuremath{\PW \to \EN}}%
\newcommand{\PWpen}{\ensuremath{\PWp \to \Pep\PGn}}%
\newcommand{\PWmen}{\ensuremath{\PWm \to \Pem\PAGn}}%
\newcommand{\Wln}{\ensuremath{\PW \to \LN}}%
\newcommand{\Zll}{\ensuremath{\Z \to \ell^+ \ell^-}}%
\newcommand{\ptZ}{\ensuremath{\pt^{\Z}}}%
\newcommand{\ptW}{\ensuremath{\pt^{\PW}}}%
\newcommand{\ptV}{\ensuremath{\pt^{\mathrm{V}}}}%
\newcommand{\x}{\ensuremath{\phantom{0}}}
\newcommand{\px}{\ensuremath{\phantom{.0}}}

\cmsNoteHeader{SMP-14-012}
\title{Measurement of the transverse momentum spectra of weak vector bosons produced in proton-proton collisions at \texorpdfstring{$\sqrt{s} = 8\TeV$}{sqrt(s) = 8 TeV}}

\date{\today}

\abstract{
The transverse momentum spectra of weak vector bosons are measured in the CMS experiment at the LHC.
The measurement uses a sample of proton-proton collisions at $\sqrt{s} = 8\TeV$, collected during a special low-luminosity running that corresponds to an integrated luminosity of $18.4\pm0.5\pbinv$.
The production of $\PW$ bosons is studied in both electron and muon decay modes,
while the production of $\Z$ bosons is studied using only the dimuon decay channel.
The ratios of $\PWm$ to $\PWp$ and $\Z$ to $\PW$ differential cross sections are also measured.
The measured differential cross sections and ratios are compared with theoretical predictions up to next-to-next leading order in QCD.
}

\hypersetup{%
pdfauthor={CMS Collaboration},%
pdftitle={Measurement of the transverse momentum spectra of weak vector bosons produced in proton-proton collisions at sqrt(s) = 8 TeV},%
pdfsubject={CMS},%
pdfkeywords={CMS, physics, Standard Model Physics, Vector boson}}

\maketitle

\section{Introduction}\label{sec:introduction}

Weak boson production processes, ${\PQq\PAQq} \to \PW + \rm{X}$ and  ${\PQq\PAQq} \to \Z / \PGg^* + \rm{X}$,
play an important role at hadron colliders.
Their clean leptonic final states allow for precise measurements with small experimental uncertainties that can be compared to theoretical predictions.

In proton-proton collisions, the $\PW$ and $\Z$ bosons (denoted as $\mathrm{V}$) are produced with zero transverse momentum $\pt$ at leading order (LO).
In a fixed-order perturbation theory, such a description shows a divergent behaviour of the $\pt$ spectrum in the low-$\pt$ region,
which is sensitive to initial-state radiation and nonperturbative effects~\cite{BalazsYuan:SoftGluonLeptonPairs}.
The high-$\pt$ region is more sensitive to perturbative effects~\cite{Melnikov:2006kv}; thus the experimental measurement of $\ptV$ constitutes a crucial test for both nonperturbative and perturbative quantum chromodynamics (QCD) calculations.

This paper reports a measurement of the $\PW$ and $\Z$ boson $\pt$ spectra and their ratios via electron and muon decay channels for the $\PW$ and the muon decay channel for the $\Z$ boson within identical lepton fiducial volumes.
The low-pileup data sample used in this analysis was collected during low instantaneous luminosity proton-proton collisions at $\sqrt{s} = 8\TeV$~\cite{IncWZ8TeVPRL}.
This sample corresponds to an integrated luminosity of 18.4\pbinv and typically has only 4 collisions per bunch crossing (pileup) resulting in less background and improved resolution compared to Ref.~\cite{CMS_ZpT8TeV}.
A finer binning at low $\Z$ boson \pt and a lower lepton \pt threshold of
20\GeV compared to the 25\GeV of Ref.~\cite{CMS_ZpT8TeV}
also provide improvements over Ref.~\cite{CMS_ZpT8TeV}.

The CDF and D0 Collaborations at the Fermilab Tevatron measured
the $\PW$ boson transverse momentum distribution in proton-antiproton collisions at $\sqrt{s} = 1.8\TeV$~\cite{TevatronWZ:CDFPhysRevLett1991,TevatronWZ:D0PhysRevLett1998}
and the inclusive $\PW$ and $\Z$ boson cross sections using the electron and muon decay channels at
$\sqrt{s} = 1.96\TeV$~\cite{TevatronWZ}.
The D0 Collaboration measured the differential cross sections of $\Z/\gamma^*$ production in the muon channel~\cite{TevatronWZ:D0PhysLettB2008} and the \pt distribution of $\Z/\gamma^*$ production in the electron or muon channel in proton-antiproton collisions at $\sqrt{s} = 1.96\TeV$~\cite{TevatronWZ:D0PhysRevLett2008_100,TevatronWZ:D0PhysLettB2010_693,TevatronWZ:D0PhysRevLett2011_106}.

The high yield of $\PW$ and $\Z$ boson events at the CERN LHC enables detailed studies of weak vector boson production mechanisms in different kinematic regions.
The ATLAS and CMS Collaborations have performed several measurements of  $\PW$ and $\Z$ boson production via leptonic decays measured at both $\sqrt{s} = 7$ and $8\TeV$.
Measurements have been made of the inclusive $\PW$ and $\Z$ boson cross sections in both electrons and muons~\cite{ATLAS7WZ,IncWZ7TeVCMS,IncWZ8TeVPRL} and of the Drell--Yan (DY) production differential cross section $\rd \sigma/\rd m$, where $m$ is dilepton invariant mass~\cite{DT7, DY8}.
The cross sections as a function of $\pt$ are measured for $\Z$ bosons~\cite{ATLAS_ZpT7TeV,ATLAS_ZptEta7TeV,CMS_ZpT7TeV,CMS_ZpT8TeV}
and $\PW$ bosons~\cite{Atlas_WpT}, but the latter has only been measured at $\sqrt{s} = 7\TeV$.
The LHCb Collaboration has measured the forward W and Z boson production cross sections and spectra for various kinematic variables
 at $\sqrt{s} = 7$ and $8\TeV$ using decays to lepton pairs
~\cite{LHCb_WZ7TeV,LHCb_Zee7TeV,LHCb_Ztau7TeV,LHCb_ZpT7TeV,LHCb_W7TeV,LHCb_WZ8TeV}.
All of the results are consistent with standard model (SM) expectations.

 The total and differential DY production cross sections are currently
 calculated up to next-to-next-to-leading-order (NNLO)
 ~\cite{Anastasiou:2003ds,Melnikov:2006kv} accuracy in perturbation
 theory, as  implemented in the \FEWZ (version 3.1) simulation code~\cite{FEWZrGavin,
   FEWZrGavinW, FEWZ_combQCD}. The theoretical treatment of soft-gluon
 emission is presently available to third order in the QCD coupling
constant using resummation techniques as used in the \textsc{ResBos} (P and CP versions)
 programs~\cite{Resbos_QCDresumBoson,Resbos_FermiTevaZboson,ResbosNPMarco}.
 The measured cross sections can also be compared with predictions from an event generator like \POWHEG (version 1.0)~\cite{POWHEG,Powheg_NLOQCD, NLOvectorPowheg, Powheg_POWHEGbOX},
which uses next-to-leading-order (NLO) QCD matrix elements.
This package uses parton shower and hadronization processes implemented in \PYTHIA (version 6.424)~\cite{Pythia64}.

The paper is organized as follows. A brief description of the CMS detector is introduced in Section~\ref{Sec:CMsDetector}.
Event samples and Monte Carlo (MC) simulations are presented in Section~\ref{sec:DataSample}.
We then describe the object reconstruction and event selection in Section~\ref{sec:EvtSelectCorr}.
These are followed by the background estimation
and the measurement of $\PW$ and $\Z$ boson $\pt$ spectra in Sections~\ref{sec:Bkg} and \ref{sec:BosonPtSpectra}, respectively.
The evaluation of the systematic uncertainties is described in Section~\ref{sec:Syst}.
We then present the results in Section~\ref{sec:Results} and the summary in Section~\ref{sec:Summary}.

\section{The CMS detector}\label{Sec:CMsDetector}

The central feature of the CMS apparatus is a superconducting solenoid of 6\unit{m} internal diameter that provides a magnetic field of 3.8\unit{T}. Within the solenoid volume are a silicon pixel and strip tracker, a lead tungstate crystal electromagnetic calorimeter (ECAL), and a brass and scintillator hadron calorimeter (HCAL), each composed of a barrel and two endcap sections. Extensive forward calorimetry complements the coverage provided by the barrel and endcap detectors. Muons are measured in gas-ionization detectors embedded in the steel flux-return yoke outside the solenoid.
A more detailed description of the CMS detector, together with definitions of the coordinate system and the relevant kinematic variables such as pseudorapidity $\eta$, can be found in Ref.~\cite{Chatrchyan:2008zzk}.

\section{Data and simulated samples}\label{sec:DataSample}

In this analysis, $\PW$ boson candidates are reconstructed from their leptonic decays
to  electrons ({$\PW \to \Pe \Pgne$}) or muons ({$\PW \to \Pgm \Pgngm$}), while
$\Z$ bosons are reconstructed only via their dimuon decays ({$\Z \to \Pgm \Pgm$}).
The candidate events
were collected by using dedicated single-lepton triggers for low instantaneous luminosity operation of the LHC that required the presence of  an electron  (muon) with
{$\pt > 22 \, (15)\GeV$} and  {$\abs{\eta}<2.5 \, (2.1)$}.

The $\PW$ and $\Z$ boson processes are generated with \POWHEG at NLO accuracy
using
the parton distribution function (PDF)
set CT10~\cite{CT10NLO}.
The factorization and the renormalization scales in the \POWHEG calculation are
set to
$(M_{\mathrm{V}}^2+(\ptV)^{2} )^{1/2}$, where $M_{\mathrm{V}}$ and $\ptV$ refer to the mass and the transverse momentum, respectively, of the vector boson.
For the background processes, parton showering and hadronization are implemented by using \PYTHIA
with the $\kt$-MLM prescription for the matrix element to parton showering matching,
as described in Ref.~\cite{Alwall:2011uj}.
For the underlying event, the Z2* tune is used.
The \PYTHIA Z2* tune is derived from the Z1 tune~\cite{Field:2010bc}, which uses the CTEQ5L PDF set, whereas Z2* adopts CTEQ6L~\cite{Puplin:Z2star}.

The effect of QED final-state radiation (FSR)
is implemented by using \PYTHIA.
The $\Z\to \Pgt \Pgt$ and diboson background event samples are generated with \PYTHIA. Inclusive $\ttbar$ and $\PW$ + jets processes are generated with the
\MADGRAPH5 (version 1.3.30)~\cite{MADGRAPH} LO matrix-element based generator package with $\mathrm{V}+n$-jets ($n = 0\ldots4$) predictions interfaced to
\PYTHIA using the CTEQ6L PDF set.
The generated events are processed through the
\GEANTfour-based~\cite{GEANT4} detector simulation, trigger emulation, and event reconstruction chain of the CMS experiment.
Independently simulated pileup events with \PYTHIA Z2* are superimposed on the generated event samples with a distribution that matches pileup events in data.

\section{Event selection}\label{sec:EvtSelectCorr}
The analysis
uses the particle-flow (PF)
algorithm~\cite{CMS-PAS-PFT-09-001,CMS-PAS-PFT-10-001}, which combines  information
from various detector subsystems  to classify reconstructed objects
or candidates  according to particle type,
thereby improving the precision of the particle energy and momentum measurements especially at low momenta.

The electron reconstruction combines electromagnetic clusters in ECAL and tracks reconstructed in the silicon tracker using the Gaussian Sum Filter algorithm (GSF)~\cite{CMS_GSF}.
Electron candidates are selected by requiring a good agreement between track and cluster variables in position and energy, as well as no significant contribution in the HCAL~\cite{Khachatryan:2015hwa}.
Electrons from photon conversions are rejected by the vertex method described in Ref.~\cite{CMS:run1_photon2015}. The magnitude of the transverse impact parameter is required to be ${<} 0.02$ cm and the longitudinal distance from the interaction vertex is required to be ${<}0.1$\unit{cm} for electrons;
this ensures that the electron candidate is consistent with a particle originating from the primary
interaction vertex,
which is the vertex with the highest $\pt^2$ sum of tracks associated to it.

The muon reconstruction starts from a candidate muon seed in the
 muon detectors followed by a global fit that uses information from the muon detectors and
the silicon tracker~\cite{CMS-PAPER-MUO-10-004}.
The track associated with each muon candidate is required to have at least one hit
in the pixel detector and at least five hits in different layers of the silicon tracker.
The track is also required to have hits in at least two different muon detector
planes.
The magnitude of the transverse impact parameter is required
to be ${<} 0.2$ cm and the longitudinal distance from the interaction vertex is required to be ${<}$0.5\unit{cm}.

The missing transverse momentum vector $\ptvecmiss$ in the event is defined as the projection of the negative vector sum of all the reconstructed particle momenta onto the plane perpendicular to the beam. Its magnitude is defined as missing transverse energy \ETmiss.

The analysis of the inclusive $\PW$ boson production in the electron (muon) channel requires events with a single isolated electron (muon) with $\pt > 25 (20)\GeV$
using the \MET distribution to evaluate the signal yield.
Background events from QCD multijet processes are suppressed by requiring isolated leptons.
For the $\PW$ boson analysis, the isolation is
 based on the particle-flow information
and is calculated by summing the \pt of
 charged hadrons and neutral particles in a cone with radius
$\Delta R = \sqrt{\smash[b]{(\Delta \eta)^{2} + (\Delta \phi)^{2}}} < 0.3\,(0.4) $ for
electron (muon) events around the direction of the lepton
at the interaction vertex
\begin{subequations}\label{eq:Iso}
\begin{align}
  I_{\rm PF}^{\Pe}&= \left( \sum  \PT^\text{charged} + \max\left[ 0, \sum \PT^\text{neutral}
               +  \sum \pt^{\gamma} - \rho  A_\text{eff}  \right] \right) /  \PT^{\Pe},\label{eq:eleIso}\\
I_{\rm PF}^{\PGm}&= \left( \sum  \PT^\text{charged} +
    {\rm max}\left[ 0, \sum \PT^\text{neutral} +  \sum \pt^{\gamma} -0.5 \sum \pt^\mathrm{PU}  \right]\right) /  \PT^{\PGm},\label{eq:muIso}
\end{align}
\end{subequations}
where $\sum  \PT^\text{charged}$ is the scalar \pt sum of charged hadrons
originating from the primary vertex, $\sum \PT^\mathrm{PU}$ is the energy deposited in the isolation cone by charged particles not associated with the primary vertex, and $\sum \PT^\text{neutral}$ and $\sum \pt^{\gamma}$ are
the
scalar sums of the \pt for neutral hadrons and photons, respectively.
A correction is included in the isolation variables to account for the neutral particles from pileup and underlying events.
For electrons, the average transverse-momentum density $\rho$ is calculated in
each event by using the ``jet area" $A_\text{jet}$~\cite{Cacciari:PUjetA}, where
 $\rho$ is
defined as the median of the $\PT^\text{jet}$$/A_\text{jet}$ distribution for all jets coming from pileup in the event,
where $\PT^\text{jet}$ is the transverse momentum of a jet.
This density is convolved with the effective area $A_\text{eff}$ of the isolation cone,
where the effective area $A_\text{eff}$ is the geometric area of the isolation cone times an $\eta$-dependent correction factor that accounts for the residual dependence of the isolation on pileup.
For muons,
the correction is applied by subtracting $\sum \PT^\mathrm{PU}$
multiplied by a factor 0.5.
This factor corresponds approximately to the ratio of neutral to charged particle production in the hadronization process.
The $\PW$ boson events are selected if
 $I_{\rm PF}^{\Pe} < 0.15$ or $  I_{\rm PF}^{\Pgm} < 0.12 $.

For the $\PW$ boson analysis, events with
a second electron with $\pt^{\Pe} > 20\GeV$
or a second muon with  $\pt^{\Pgm} > 17\GeV$ that passes loose selection criteria are rejected as $\PW$ boson events
to reduce the background contributions from the $\Z/\PGg^*$ DY processes.
The second electron selection uses a loose selection working point~\cite{Khachatryan:2015hwa}, which mainly relaxes the match of the energy and position between the GSF tracks and the associated clusters in the ECAL.
For the second muon, the required number of hits in the pixel detector, the silicon tracker, and the muon detector are relaxed~\cite{CMS-PAPER-MUO-10-004}.

Several corrections  are applied to the simulated events to account for
the observed small discrepancies between data and simulation.
A better description of the data is obtained
by applying corrections to the lepton \pt
and \MET.
There are two main sources of disagreement in the \pt description:
the momentum scale and the modeling of the \pt resolution.
The corrections for these effects
are determined from a comparison of the $\Zll$ mass spectrum between data and simulation~\cite{IncWZ7TeVCMS}.
The lepton momentum scale correction factor is found to be  close to unity
with an uncertainty of   0.2\% (0.1\%) for electrons (muons).
An additional smearing of the lepton $\pt$- and $\eta$-dependent resolution in the range 0.4 to 0.9 (0.1 to 0.7)\GeV for electrons (muons) is
applied to reproduce the distribution of the dilepton invariant mass observed in data.

The vector boson recoil is defined as the vector sum of the transverse momenta
of all the observed particles, excluding the leptons produced in the vector boson decay.
The \MET spectra in the $\PW$ boson signal simulation rely on the modeling of the $\PW$ boson recoil and
the simulation of the detector response.
The correction factors for the $\PW$ boson recoil simulation are estimated using a comparison of the $\Z$ boson recoil between data and simulation~\cite{CMS:met7_TeV,IncWZ7TeVCMS}.
The factors for the recoil scale (resolution) range from 0.88 to 0.98 (from 0.84 to 1.09) as a function of the boson \pt with an uncertainty of about 3 (5)\%. They are applied to the simulated $\PW$ boson recoil distributions.

The corrected \MET and corrected lepton momenta are used to calculate
the transverse mass $\MT$ of the $\PW$,
\begin{equation}\label{transMass}
\MT = \sqrt{2 \, \pt^\ell \,  \MET(1-\cos \Delta\phi_{\MET, \ell})},
\end{equation}
where $\Delta\phi_{\MET, \ell}$ is the azimuthal angle between \ptvecmiss and lepton $\vec{\pt}$.
 $\MT$~is used for the signal yield extraction for the muon channel
in the high-\pt region, as described in Section~\ref{subsec:WsigExt}.

A set of lepton efficiencies, namely the lepton reconstruction and identification, and trigger efficiencies, are estimated in simulation and then corrected for the differences between data and simulation.
These corrections are evaluated by using a ``tag-and-probe'' method~\cite{CMS:2011aa} and the total efficiency correction factor for the simulated samples ranges between $0.92\pm0.03$ ($0.93\pm0.05$)  and $1.03\pm0.08$ ($1.04\pm0.03$) for electrons (muons).

For the inclusive $\Z$ boson events we require two isolated oppositely charged muons with $\pt > 20\GeV$.
A vertex fit is performed to ensure that the candidates originate from the same $\Z$ boson.
The background due to cosmic ray muons passing through the detector and mimicking dimuon events is
suppressed by
requiring that the two muons are not back-to-back, \ie the three-dimensional
opening angle between the two muons should be smaller than $\pi - 0.02$ radians. Finally, the muon pair is required to have a
reconstructed invariant mass in the range 60--120\GeV.

For the $\Z$ boson analysis, the dimuon invariant mass selection and a vertex fit enables the use of a simpler isolation variable based only on charged tracks.
The track isolation variable $I_{\rm trk}$ is defined as the
scalar sum of the track momenta of charged particles lying within a cone of radius
$\Delta R =0.3$ around the muon direction.
The muons are isolated if $I_{\rm trk}/\pt^\Pgm < 0.1$.

\section{Measurement of the transverse momentum spectra}\label{sec:BosonPtSpectra}

The transverse momentum of the vector boson $\ptV$ is computed from the momentum sum of the decay leptons for the $\Z$ boson, or the lepton and $\ptvecmiss$ for the $\PW$ boson.
The measurements are performed within the lepton fiducial volumes defined by {$\pt > 25 \, (20)\GeV$}, {$\abs{\eta} < 2.5 \, (2.1)$} for the electron (muon) channel.
The fiducial region for the boson differential cross section is defined by the \pt and $\eta$ requirements on the leptons.

The transverse momentum spectra are analyzed as binned histograms, with bin widths  varying from 7.5 (2.5)\GeV for the $\PW$ ($\Z$) boson up to 350\GeV,
in order to provide sufficient resolution to observe the shape of the
distribution, limit the migration of events between neighbouring bins,
and ensure a sufficient number of events in each bin.
The cross section in the $i$th $\ptV$ bin is defined as
\begin{equation}\label{eq:WpT}
  \frac{\rd\sigma_{i}}{\rd p^{\mathrm{V}}_{{\rm T},\, i}} = \frac{N_i}{\Delta_i \, \epsilon_i  \int L \rd t},
\end{equation}
where $N_i$ is the estimated number of signal events in the bin, $\Delta_i$ is the width of the bin,
$\epsilon_i$ is the efficiency of the event selection in that bin,
and $\int L \rd t$ is the integrated luminosity.

The differential distributions are unfolded to the lepton level before QED final-state radiation (pre-FSR) within the same fiducial volume.

\subsection{The \texorpdfstring{$\PW$}{W} boson signal extraction}\label{subsec:WsigExt}

\begin{figure}
\begin{center}
\includegraphics[width=0.48\textwidth]{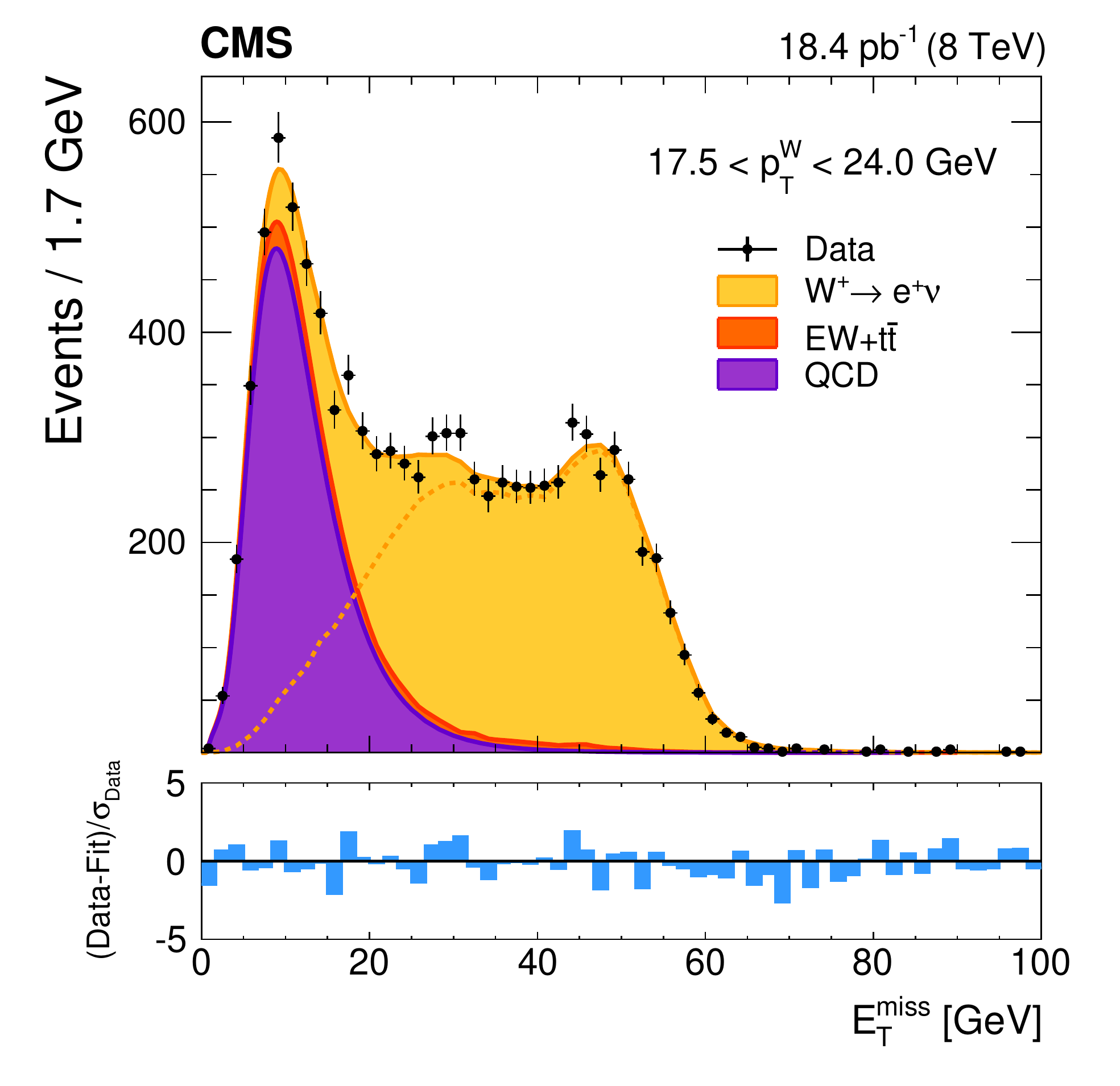}
\includegraphics[width=0.48\textwidth]{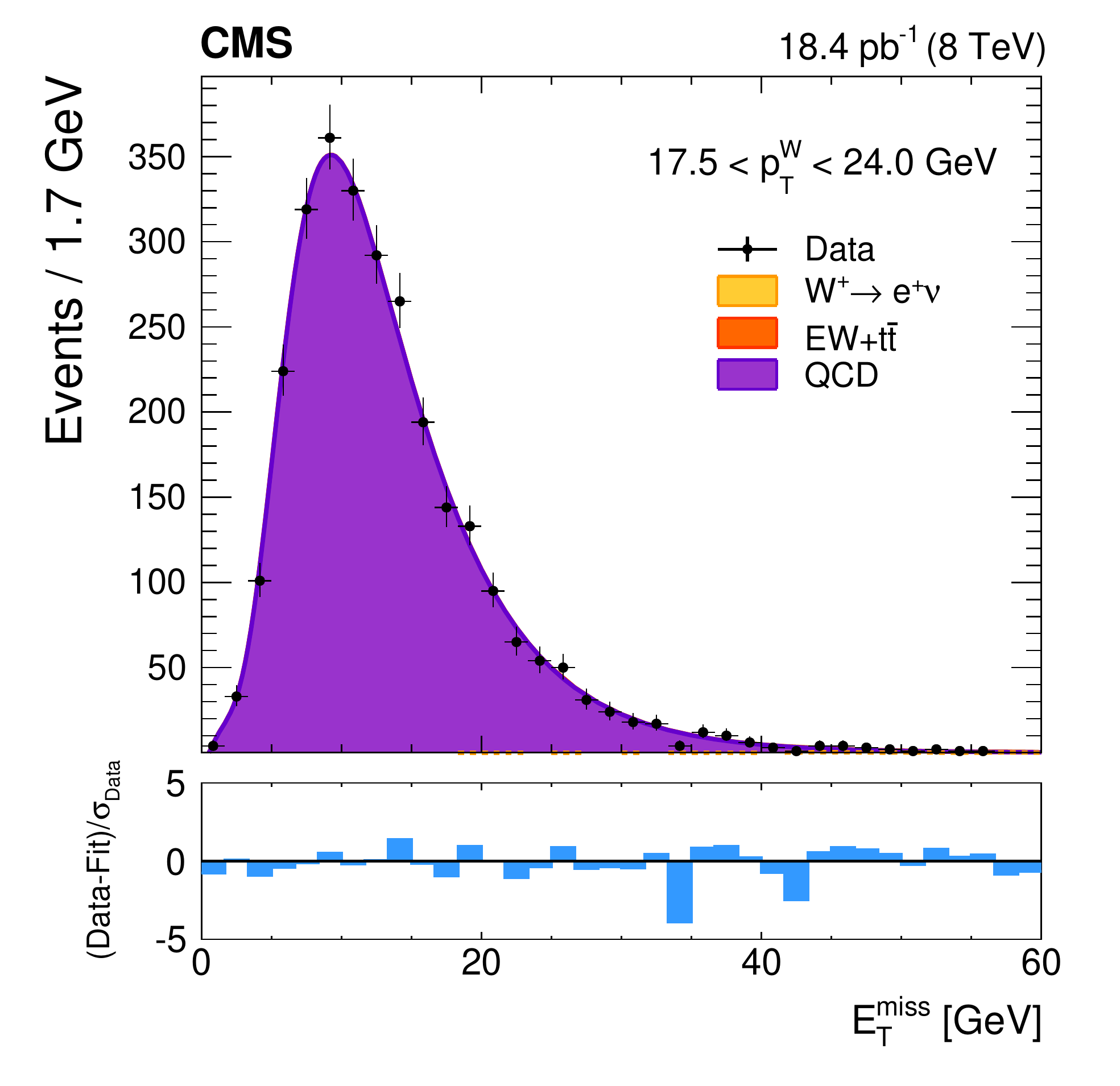}
\includegraphics[width=0.48\textwidth]{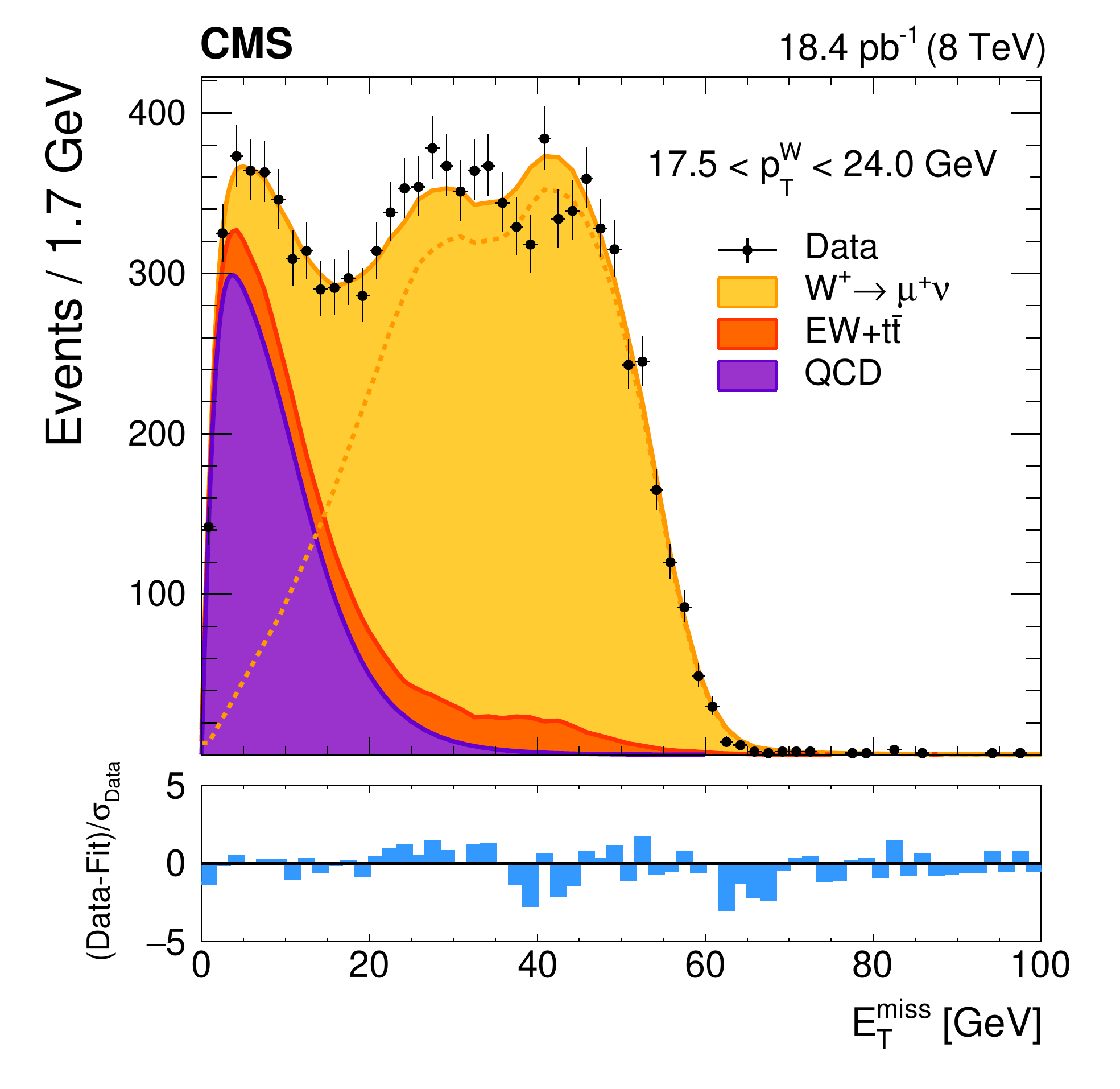}
\includegraphics[width=0.48\textwidth]{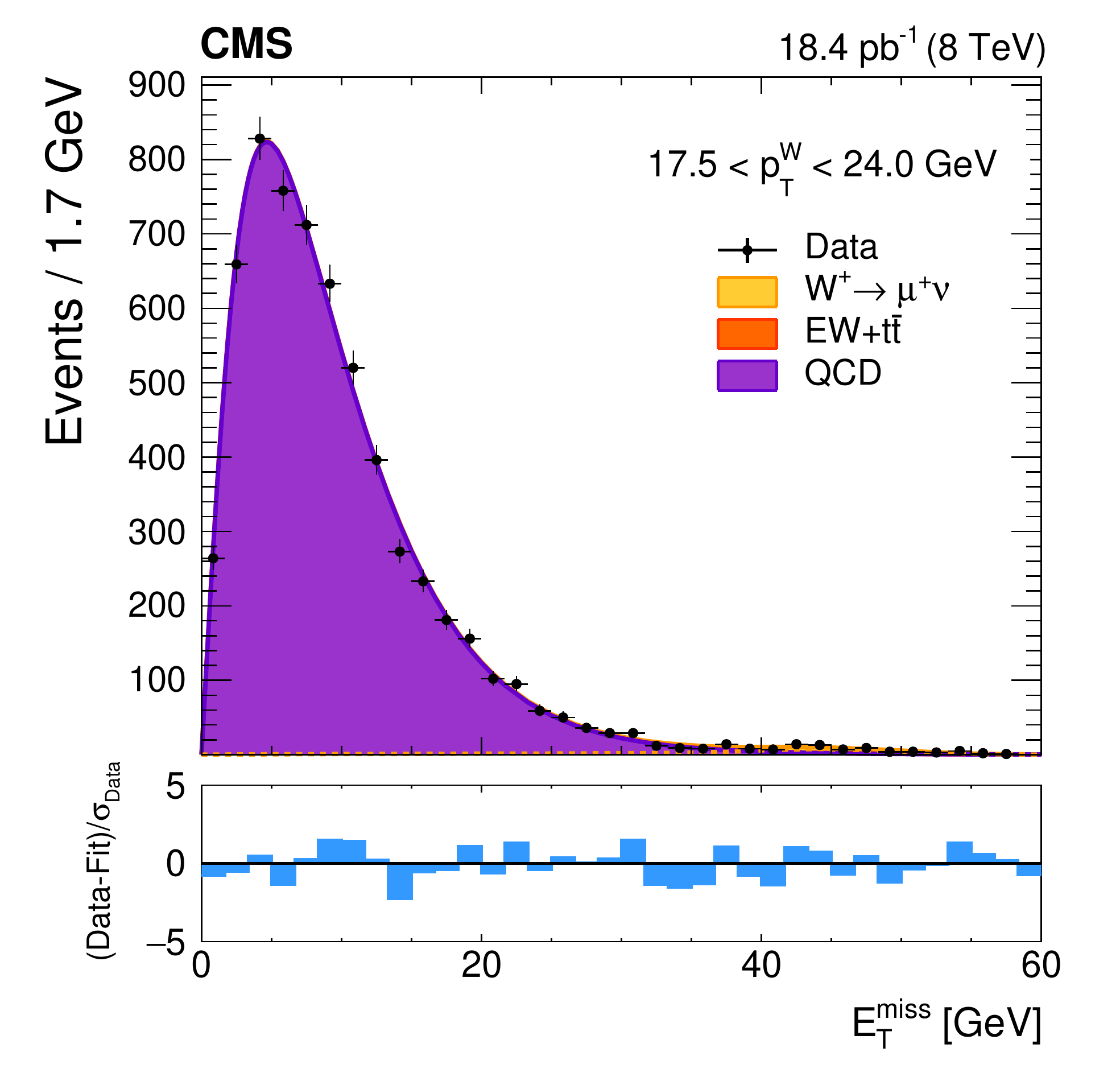}
\caption{\label{fig:sigFitExample}
The \MET distributions for the selected \PWpen~(upper) and \PWpmn~(lower) candidates for $ 17.5 < \ptW < 24$\GeV (left)
and the corresponding QCD multijet-enriched control sample (right).
Solid lines represent the results of the fit.
The dotted lines represent the signal shape after background subtraction.
The bottom panels show the difference between data and fitted results divided by the statistical uncertainty in data, $\sigma_\text{Data}$.
}
\end{center}
\end{figure}

The W boson signal yield and the backgrounds for each $\ptW$ bin are determined using an extended likelihood fit to the \MET distributions.
The fits constrain the sum of signal plus background to the data within each bin.
Fig.~\ref{fig:sigFitExample} shows an example of the fit for the bin  $17.5 < \ptW < 24$\GeV.
The signal and background shapes are determined separately for \PWp ~and \PWm ~bosons to account for the difference in the kinematical configuration arising from the parity-violating nature of the weak interaction.
The signal yield and background contaminations are estimated from the fit,
which is performed simultaneously in the signal candidate sample and in the corresponding QCD control sample for each $\ptW$ bin.
The QCD multijet-enriched control samples are defined
by inverting the selection on some
identification variables for the electron channel, and by inverting
 the isolation requirement for the muon channel,
while maintaining the rest of the signal selection criteria.

The $\PW$ boson signal and electroweak (EW) background (explained in Section~\ref{sec:Bkg}) templates are produced by using simulated events
 including all corrections described in Section~\ref{sec:EvtSelectCorr}.
The EW contribution is constrained for the $\PW$ signal yield by fixing the ratio of the theoretical cross section
of the EW contribution to that of $\PW$ boson production.
The QCD shape of \MET distribution is parameterized
by a modified Rayleigh function~\cite{IncWZ8TeVPRL},
\begin{equation}\label{eq:qcdPdf}
  f(x) = x \exp \left\lgroup -\frac{x^2}{2(\sigma_0+\sigma_1 x)^2} \right\rgroup,
\end{equation}
where $\sigma_0$ and $\sigma_1$ are free parameters of the fit.
The fit uses $x = \MET$ for $\ptW > 17.5\GeV$
 and $x = (\MET - a) $ for $\ptW < 17.5\GeV$, where $a$
is a parameter of the fit needed to take into account
the minimum \MET value at each $\ptW$ bin due to trigger
requirements on the $\pt^\ell$.
The parameter $\sigma_0$ in Eq.(\ref{eq:qcdPdf}) is, however, kept floating separately in signal and control regions.

\begin{figure}
\begin{center}
\includegraphics[width=0.48\textwidth]{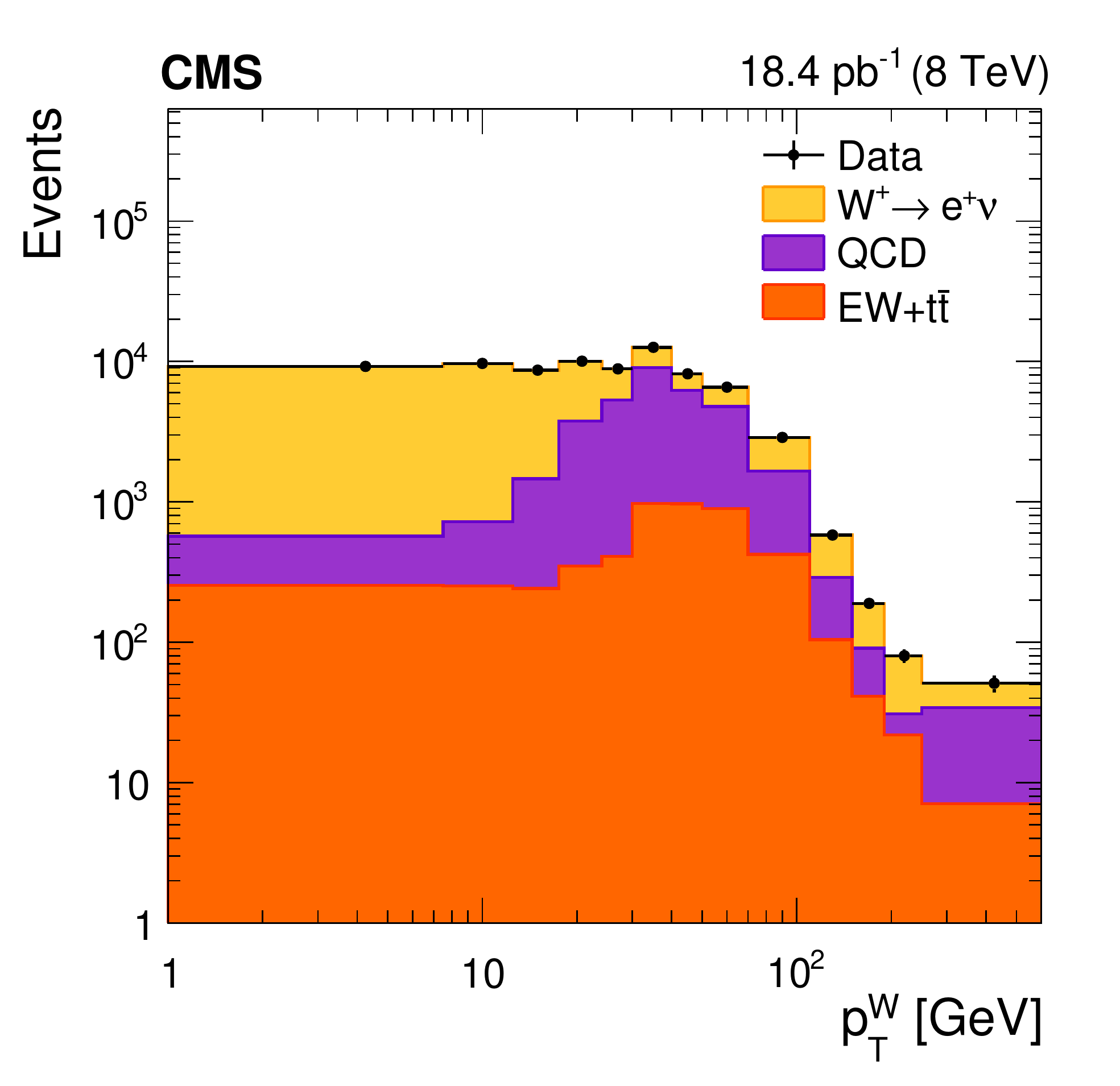}
\includegraphics[width=0.48\textwidth]{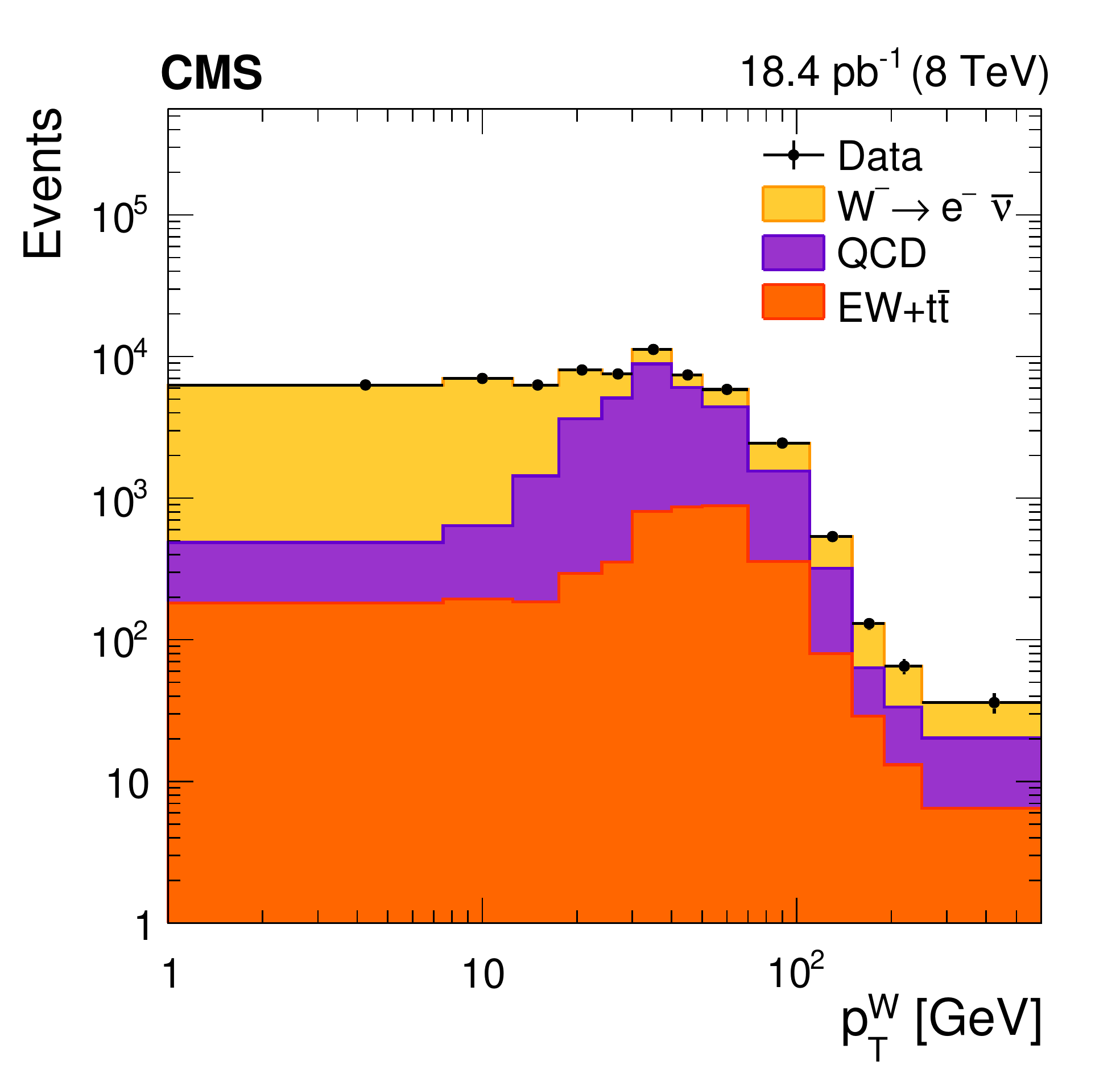}
\includegraphics[width=0.48\textwidth]{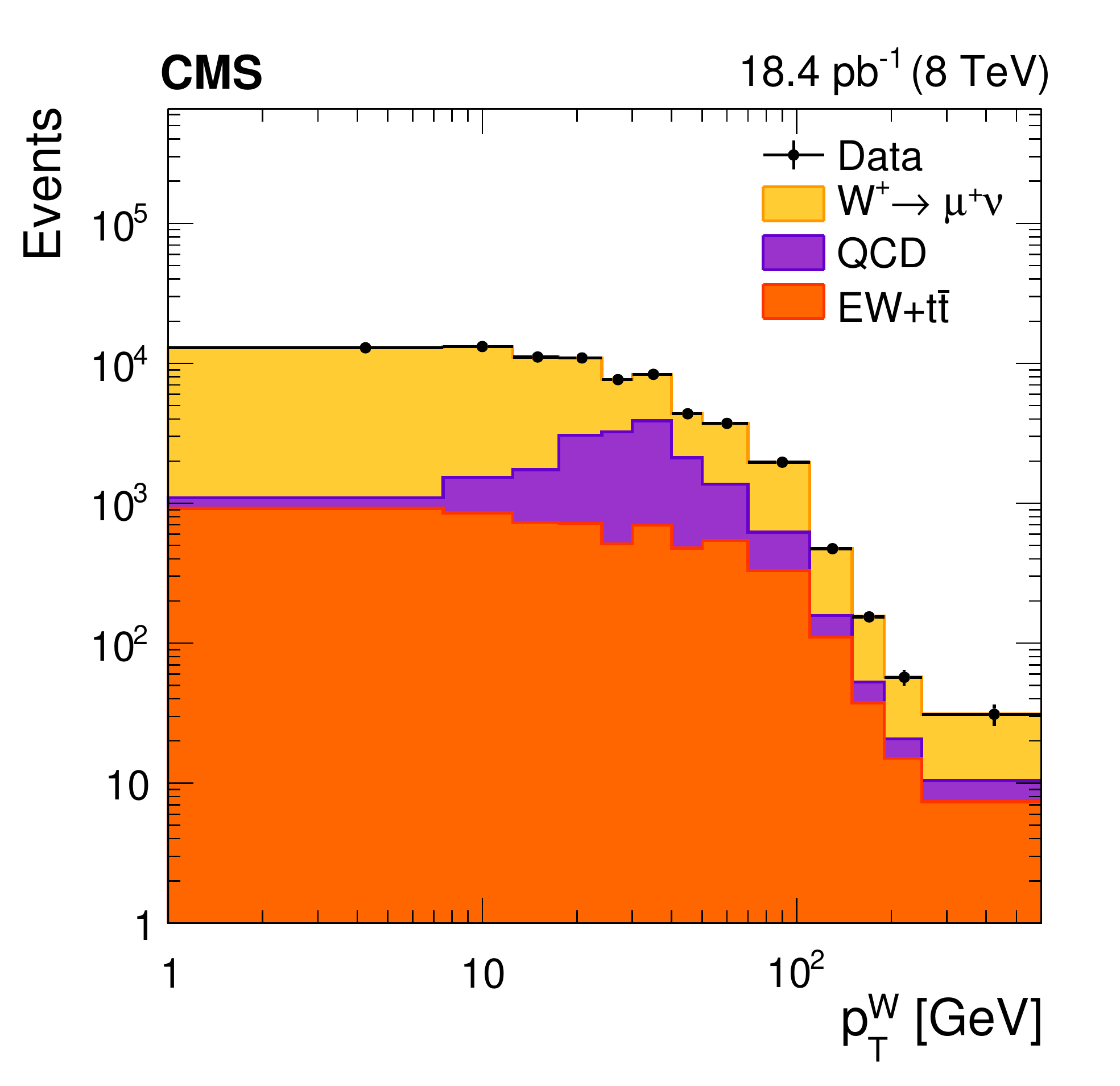}
\includegraphics[width=0.48\textwidth]{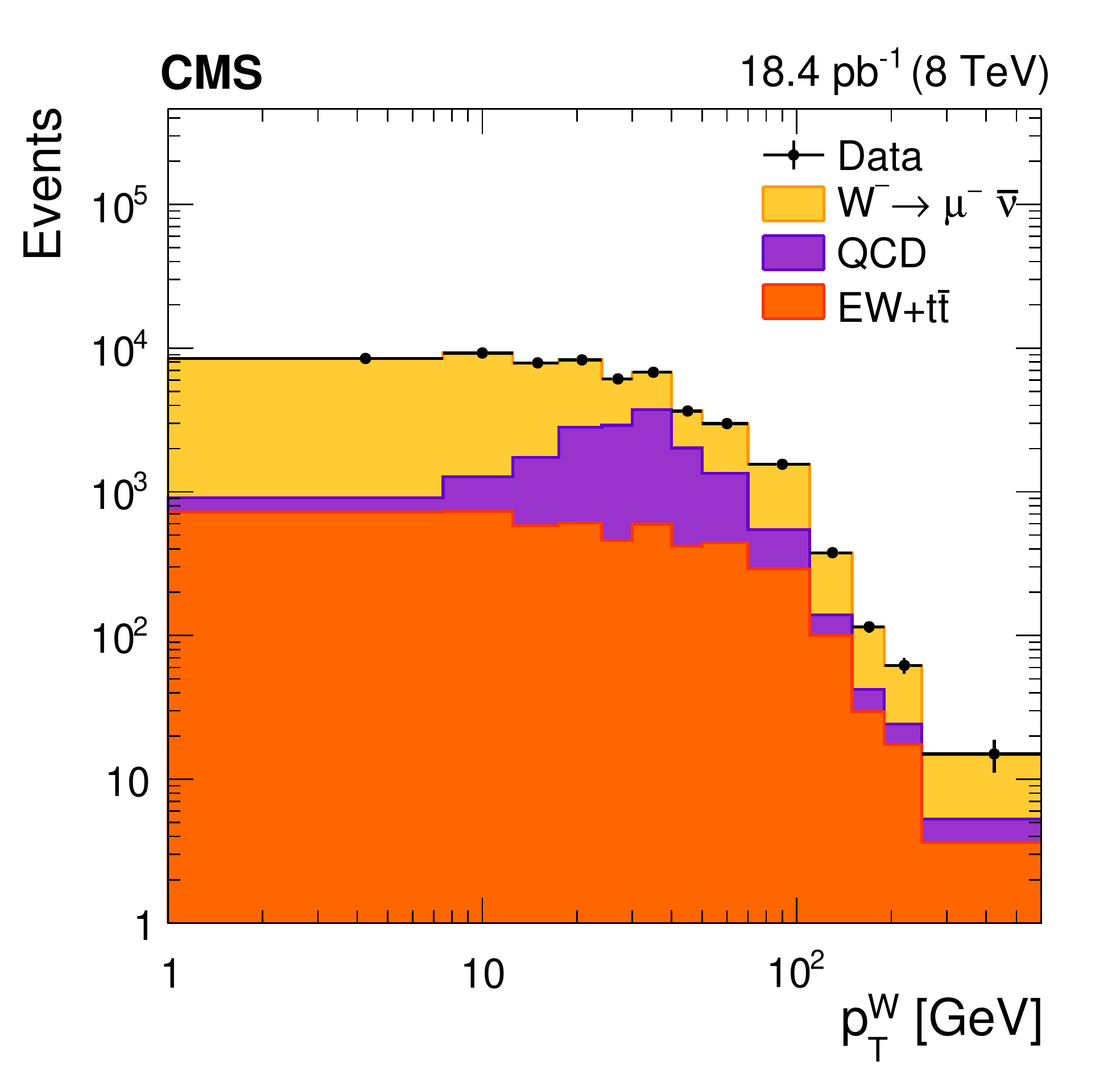}
\caption{\label{fig:FitWDistribution}
Signal and background yields after fitting the data for \PWpen~(upper left), \PWmen~(upper right), \PWpmn~(lower left), and \PWmmn~(lower right) as a function of the $\PW$ boson \pt.
The points are data yields with statistical uncertainties.
The stacked histogram shows the signal and background components estimated from a fit to the \MET or \MT~ distribution at each $\PW$ boson \pt bin.
}
\end{center}
\end{figure}

In the muon channel, the QCD multijet contribution decreases noticeably with increasing $\ptW$ because the probability of the background muon to pass the  isolation criteria decreases.
For $\ptW > 70\GeV$ the $\MT$ distributions, instead of \MET, are fitted to maintain a good separation between the signal and the QCD background shape.
The extracted signal and background yields are shown as a function of $\ptW$
 in Fig.~\ref{fig:FitWDistribution} for electrons (upper) and muons (lower).

In order to obtain the differential cross section before FSR, the detector resolution and FSR effects need to be corrected.
This is achieved by a two-step unfolding process using the singular value decomposition (SVD) method~\cite{SVD_NuclInstrumMeth}. SVD uses two response matrices.
The first matrix maps the intra-bin migration effects to the reconstructed $\ptW$ from leptons after a possible FSR (post-FSR) effect, using the \POWHEG simulated signal sample as the baseline, after applying lepton momentum resolution, efficiency, and recoil corrections.
The second matrix maps the $\ptW$ distribution taking into account the FSR effect of the lepton, \ie from pre-FSR to post-FSR.

The event reconstruction efficiency is corrected bin-by-bin after unfolding for the detector resolution by using the simulated signal sample.
An acceptance correction is applied to the pre-FSR distribution after FSR unfolding;  about 5.1\% (1.9\%) of the events with a pre-FSR level electron (muon) generated within the fiducial region do not pass the post-FSR lepton requirements of the fiducial volume.

\subsection{The \texorpdfstring{$\Z$}{Z} boson signal extraction}

\begin{figure}[hbtp]
  \begin{center}
    \includegraphics[width=0.48\textwidth]{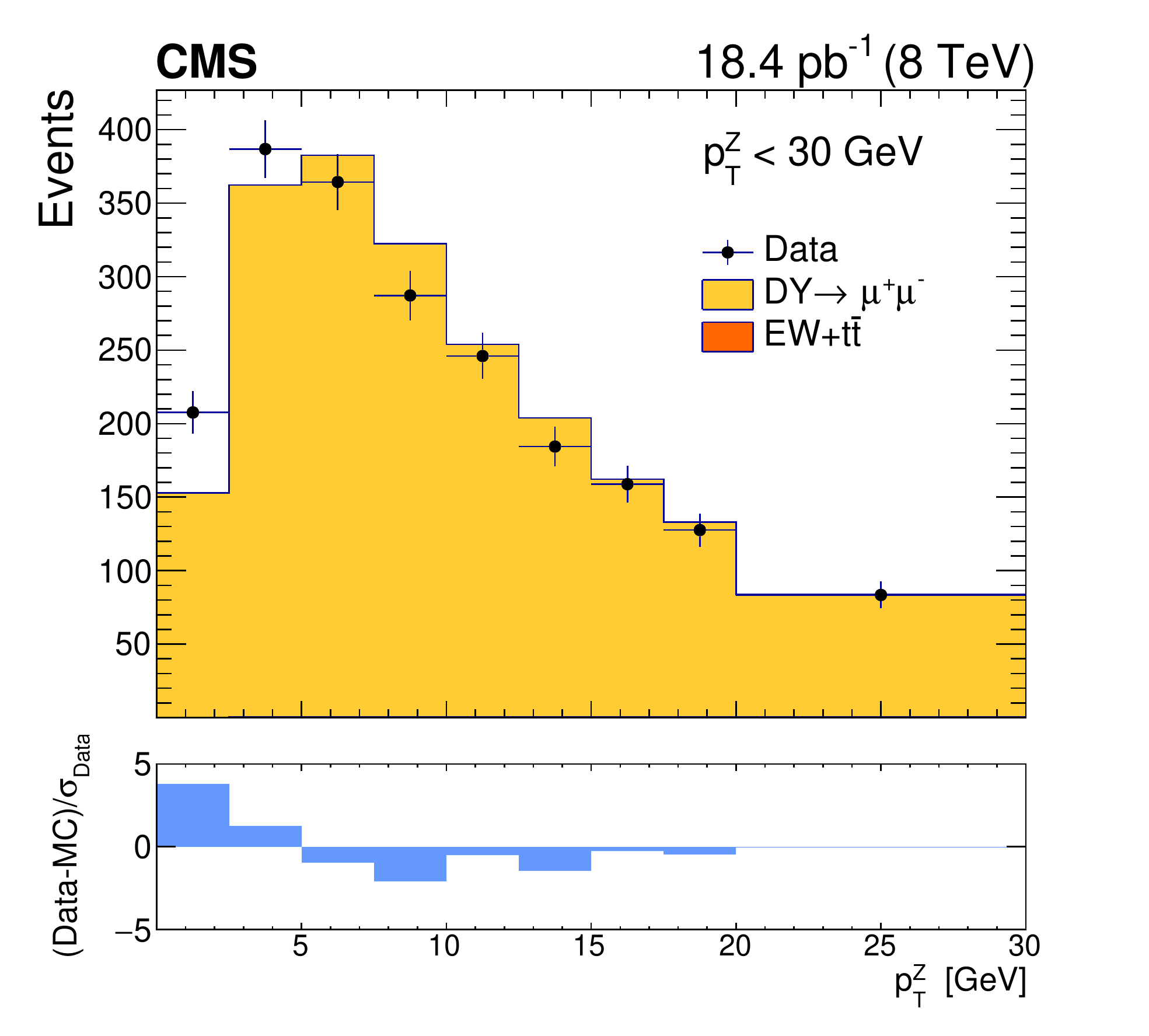}
	\includegraphics[width=0.48\textwidth]{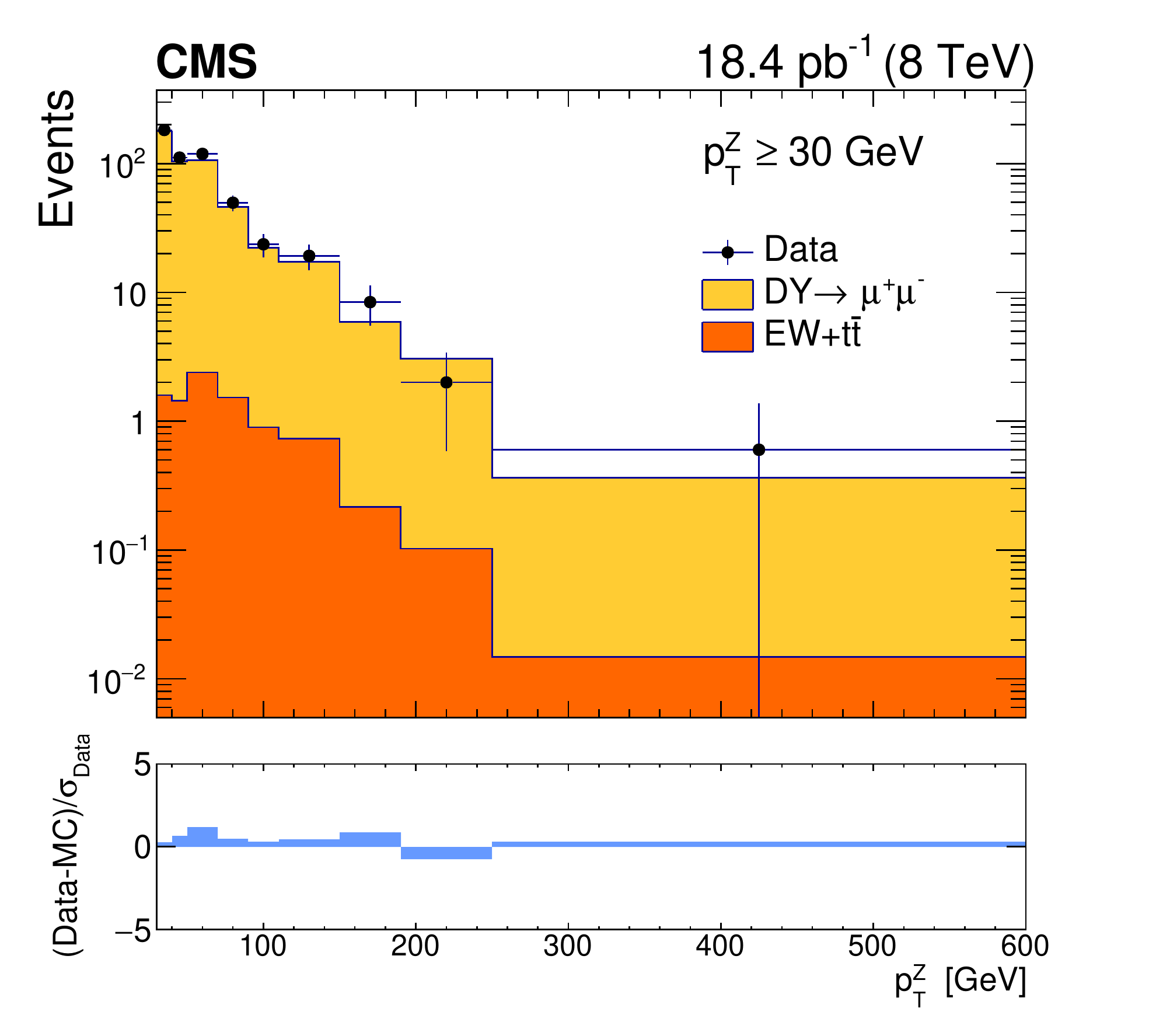}
    \caption{
	     Data and simulated events for both
             DY processes and various backgrounds after event reconstruction.
	     Left (right): events for low (high) $\ptZ$, {$\ptZ< 30~(\geq 30)\GeV$}.
             The lower panels show the difference between the data and the simulation predictions divided by the statistical uncertainty in data, $\sigma_\text{Data}$.}
    \label{fig:zPt}
  \end{center}
\end{figure}

The number of observed $\Z$ boson events
is obtained by subtracting  the estimated number of background events from the total number of detected events in each of the $\ptZ$ bins.
The  transverse momentum distribution of the dimuon system for
the reconstructed events is shown in Fig.~\ref{fig:zPt} separately for the low- and high-$\ptZ$ regions to show the level of agreement between data and simulation.
The NLO QCD calculation in \POWHEG underestimates the data by 27\% in the  $\ptZ$ range below 2.5\GeV.

The measured $\ptZ$ distributions are
corrected for bin migration effects that arise from the detector resolution and FSR effects with a similar technique to the $\PW$ boson analysis described in Section~\ref{subsec:WsigExt} using a matrix-based unfolding procedure~\cite{bib:unfold2}.
The final result is corrected by the bin width and is
normalized by the measured total cross section $\sigma$ within the fiducial region (Section~\ref{sec:BosonPtSpectra})  in the range of the dimuon mass,  {$60 < m_{\PGm\PGm} < 120\GeV$}.

\section{Background estimation}\label{sec:Bkg}
\subsection{The \texorpdfstring{$\PW$}{W} boson analysis}

QCD multijet events are the dominant source of background in the $\PW$ boson analysis.
The level of contamination is estimated from data as described in Section~\ref{subsec:WsigExt}.
It is about 40\% and 19\% of the selected \Wen~and \PWmn~event yields, respectively.

The contributions of EW and $\ttbar$ background sources are estimated by using simulated events.
The DY processes with  $\Z/\PGg^*\to \ell^+ \ell^-$~contribute to the \Wln~background  when
one of the two leptons is not detected. These processes account for
approximately  4.7\% (5.0\%) of the selected events in the electron (muon) channel.
Events from \Wtn~(where the  $\Pgt$ decays leptonically)
have, in general, a softer lepton than the signal events. They are strongly
suppressed by using a high value of the  minimum $\pt^{\Pe,\Pgm}$ requirement
for acceptance. The background contribution from \Wtn~ is 1.7\% (3.3\%)
of selected events in the electron (muon) channel.
The background originating from $\ttbar$ production is estimated to be
0.35\% (0.41\%) of the selected events, while that from  boson pair production ($\PW\PW$, $\PW\Z$, and $\Z\Z$)
is even smaller, about  0.03\% of the selected events for both decay channels.

\subsection{The \texorpdfstring{$\Z$}{Z} boson analysis}
The main sources of background in the dimuon analysis are
$\Z\to\PGt\PGt$, \ttbar, $\PW$+jets, and diboson ($\PW\PW$, $\PW\Z$, and $\Z\Z$) production
with the subsequent decay of $\PW$, $\Z$, and $\PGt$ to muons.
The simulation of these backgrounds is validated with data
by measuring the \pt
of the final state with an electron and a muon.
The residual background contribution is due to QCD multijet hadronic processes that contain energetic muons,
predominantly from the semileptonic decays of B hadrons.
A control sample of events with a single muon that passes all the requirements of this analysis except the isolation criteria is selected to estimate the contribution of this source.
This sample is subsequently used to estimate the probability for a
muon to pass the isolation
requirements as a function of the muon $\pt$ and $\eta$.
This probability  is used to predict the number of background events
with two isolated muons based on a sample of events with two
nonisolated muons. This procedure, which is validated by using
simulated events, predicts a negligible contribution from QCD multijet
production over the full range of our $\ptZ$ spectrum.
After the full selection, the background contamination, which consists primarily of $\Z\to \PGt\PGt$ and \ttbar~processes, with an uncertainty dominated by the statistical uncertainties in the background simulation is estimated to be less than 1\%  of the total event yield.

\section{Systematic uncertainty}\label{sec:Syst}

The leading  sources of systematic uncertainties are mostly common to both the $\PW$ and $\Z$ boson analyses.
They include the determination of the correction factors for the lepton efficiency (reconstruction, isolation, and trigger),
the electron or muon momentum resolution parameters,
and the construction of the response matrices for unfolding the detector resolution and FSR effects.
The simulated distributions are corrected for the efficiency differences between data and simulation using scale factors obtained from the tag-and-probe method.
The variation of the measured scale factors due to different choices of
signal and background models and the
\pt and $\eta$ binnings for the measured lepton
 are treated as systematic uncertainties.
The momentum resolution is estimated by comparing data and the simulated $\Z$ boson mass distribution.
The uncertainties in the parameterization of the mass distribution are propagated in the resolution calculation.
The uncertainty in the model-dependent FSR simulation is estimated by reweighting the simulated data samples.
We are using event-dependent weights from a soft collinear approach~\cite{FSRsoft} and higher-order corrections in $\alpha(\pt^2)$~\cite{HadronContQED}.
The difference in signal yields before and after reweighting is
assigned as a systematic uncertainty.
The systematic uncertainty in the luminosity measurement is completely canceled out
since the results are presented as normalized distributions.

The uncertainty in the recoil corrections to \MET is taken into account for the $\PW$ boson analysis.
The systematic uncertainty associated with the shape of the \MET distribution from the QCD multijet process is estimated by
introducing an additional term $\sigma_2  x^2$ into Eq.(\ref{eq:qcdPdf}), where $\sigma_2$ is another shape parameter to describe the tail of \MET at the second order, and repeating the fit procedure.
A set of pseudo-experiments is generated by varying all parameters of the equation within their uncertainties.
The bias in the measured values with the pseudo-experiments provides the systematic uncertainty in the parameterization of the shape.
An additional uncertainty is assigned due to the simultaneous fit procedure by
floating the tail parameter $\sigma_1$ in the extraction of the signal yields.
These are used to estimate the shape dependence of the fits to the QCD multijet-enriched control samples.

The cross section for each of the EW backgrounds in the $\PW$ boson analysis
is varied around the central value within its uncertainty
and the resulting fluctuation of signal yield extraction by the fit in each $\ptW$ bin
is assigned as a systematic uncertainty.

The unfolding procedure is sensitive to the statistical uncertainties in the construction of the response matrix.
These uncertainties range from 0.1\% to 1.0\% depending on the channel and  $\ptV$ bin.
The boson distributions are compared with those obtained by using an alternative response matrix derived from a different generator, \MADGRAPH5.
The difference is taken as the unfolding bias.

The background for the dimuon final state is measured from simulation with
correction factors derived from data, the corresponding uncertainty is estimated
 by varying its contribution.
The uncertainty is about 0.4\% level up to $40\GeV$ of dimuon $\pt$.

\section{Results}\label{sec:Results}
The fiducial cross sections at pre-FSR level
are calculated as the sum of contributions from all bins and listed in Table~\ref{tab:totCross}.
\begin{table}[htb]
\topcaption{
The fiducial cross sections at pre-FSR level calculated as the sum of differential cross sections.
The fiducial volumes are defined in Section~\ref{sec:BosonPtSpectra}.
}
\begin{center}
\begin{tabular}{c|c}
\hline
Channel                             & $\sigma \, \mathcal{B}$ [nb] (fiducial) \\
\hline
$\cPZ\to \mu^+\mu^-$        & $0.44\pm0.01\stat\pm0.01\syst\pm0.01\lum$ \\
$\rm{W}$ $\to {\rm{e}}\nu$  & $6.27\pm0.03\stat\pm0.10\syst\pm0.16\lum$\\
$\rm{W}$ $\to \mu\nu$       & $6.29\pm0.02\stat\pm0.09\syst\pm0.16\lum$ \\
\hline
\end{tabular}
\label{tab:totCross}
\end{center}
\end{table}

The low-pileup data is adjusted to the lepton fiducial volume at post-FSR level used in Ref.~\cite{IncWZ8TeVPRL}.
The results are $0.40\pm0.01\stat\pm0.01\syst\pm0.01\lum$\unit{nb} for the $\cPZ$ channel and
$5.47\pm0.02\stat\pm0.06\syst\pm0.14\lum$\unit{nb} for the mean value of $\PW$ electron and muon channel results weighted by uncertainties.
These are consistent with the supplemental material of Ref.~\cite{IncWZ8TeVPRL}, where the fiducial inclusive $\cPZ$ boson cross section is $0.40\pm0.01\stat\pm0.01\syst\pm0.01\lum$\unit{nb} and the $\PW$ boson cross section is $5.42\pm0.02\stat\pm0.06\syst\pm0.14\lum$\unit{nb}.

The differential cross sections $\rd\sigma/\rd\ptV$, corrected for FSR, are normalized to the total fiducial cross section.
Some uncertainties are canceled in the normalized cross sections,
thus allowing for a more precise shape comparison.
The uncertainties in the measurement of the lepton efficiencies are decreased by factors of 1.6 to 7.7 with respect to the cross section before the normalization.
The uncertainties in the EW background cross sections affect both the numerator and the denominator, hence the corresponding uncertainty is decreased by a factor of 20.
The other sources of uncertainty remain at a level similar to the differential cross section measurements before normalization.

The differential cross sections in the electron and muon channels, derived individually for $\PWp$ and $\PWm$ bosons,  are combined after taking into account the possible correlations.
The systematic uncertainties due to FSR and EW background cross sections are added linearly
under the assumption that these uncertainties are 100\% correlated.
All other charge-dependent uncertainties are assumed to be uncorrelated and are added in quadrature.

The data unfolded to the pre-FSR level are compared to various theoretical predictions:
\textsc{ResBos}-P version (CP version) with scale (scale and PDF) variation for the $\PW$ ($\Z$) boson result, \POWHEG with PDF uncertainty,
and \FEWZ with PDF and renormalization and factorization scale uncertainties.
\textsc{ResBos} adopts the Collins--Soper--Sterman formalism with four parameters (C1, C2, C3, and C4)
for the resummation of the multiple and collinear gluon emissions~\cite{CollinsBackToBackJet, CollinsPt},
which yields a next-to-next-to-leading-order accuracy.
It allows also for the use of a $K$ factor grid to get an effective NNLO description.
The scale parameters  in C2 ($\mu_F$) and C4 (for $\alpha_s$ and PDF) are set to
$M_{\ell\ell}/2$ (where $M_{\ell\ell}$ is the invariant mass of the lepton pair) as the nominal value and
different grid points are generated with scale variations $M_{\ell\ell}$ and $M_{\ell\ell}/4$ for the determination of the
scale uncertainty.
The nonperturbative function implemented in \textsc{ResBos} affects mostly the low-\pt region around 1--4\GeV
and the intermediate-\pt region with small contribution.

\subsection{The \texorpdfstring{$\PW$ and $\Z$}{W and Z} differential cross sections}
The numerical results and all of the uncertainties for the normalized differential cross section are listed in Tables~\ref{tab:ElectronSyst} and~\ref{tab:MuonSyst}
for the electron and muon channels of the $\PW$ boson decay, respectively.
The results for the $\ptZ$ spectrum are summarized in Table~\ref{tab:ZMNormSystFid}.
After combining the effects discussed in Section~\ref{sec:Syst}, the total systematic uncertainty in each
bin is found to be smaller than the corresponding statistical uncertainty for the $\Z$ boson and at a similar level for the $\PW$ boson except in the high-$\ptW$ region.

\begin{table}[htb]
\renewcommand\arraystretch{1.2}
\topcaption{
The $\PW$ boson normalized differential cross sections for the electron channel in bins of $\ptW$, (1/$\sigma$)(d$\sigma$/d\pt) ({$\PW \to \Pe\PGn$}),
and systematic uncertainties from various sources in units of~\%,
where $\sigma$ is the sum of the cross sections for the $\ptW$ bins.
(1/$\sigma$)(d$\sigma$/d\pt) is shown with total uncertainty, \ie the sum of statistical and systematic uncertainties in quadrature.
}
\begin{center}
\resizebox{\textwidth}{!}{
\begin{tabular}{c|c|c|c|c|c|c|c|c|c|c|c|c}
\hline
\multirow{1}{*}{Bin}   &\multirow{1}{*}{Lept.} &\multirow{1}{*}{Mom.}&\multirow{1}{*}{\MET}&\multirow{1}{*}{QCD}  &\multirow{1}{*}{QCD}  &\multirow{2}{*}{EW}&\multirow{1}{*}{SVD} &\multirow{2}{*}{FSR}&\multirow{1}{*}{Unfld.}&\multirow{1}{*}{Total}&\multirow{2}{*}{Stat.}&\multirow{1}{*}{(1/$\sigma$)(d$\sigma$/d\pt)}\\
\multirow{1}{*}{(\GeV)}&\multirow{1}{*}{recon.}&\multirow{1}{*}{res.}&\multirow{1}{*}{res.}&\multirow{1}{*}{bkgr.}&\multirow{1}{*}{shape}&   &\multirow{1}{*}{unfld.}&   &\multirow{1}{*}{bias}&\multirow{1}{*}{syst.}& &\multirow{1}{*}{($\GeV^{-1}$)} \\
\hline
\multirow{1}{*}{\px0--7.5}     & 0.31 & 0.21 & 0.22 & 0.51 & 0.20 & 0.05 & 0.08 & 0.05 & 0.75 & 1.03 & 0.60  & (4.74 $\pm$ 0.06) $\times$ 10$^{-2}$ \\
\multirow{1}{*}{\x7.5--12.5}    & 0.26 & 0.09 & 0.10 & 0.64 & 0.26 & 0.04 & 0.08 & 0.05 & 1.43 & 1.62 & 0.74  & (4.12 $\pm$ 0.07) $\times$ 10$^{-2}$ \\
\multirow{1}{*}{12.5--17.5}   & 0.17 & 0.24 & 0.10 & 0.48 & 0.37 & 0.02 & 0.08 & 0.04 & 1.11 & 1.31 & 0.89  & (2.42 $\pm$ 0.04) $\times$ 10$^{-2}$ \\
\multirow{1}{*}{17.5--24\px}   & 0.16 & 0.30 & 0.27 & 0.66 & 0.43 & 0.04 & 0.09 & 0.00 & 0.36 & 0.98 & 0.95  & (1.49 $\pm$ 0.02) $\times$ 10$^{-2}$ \\
\multirow{1}{*}{24--30}   & 0.37 & 0.26 & 0.35 & 0.80 & 0.51 & 0.05 & 0.10 & 0.06 & 0.58 & 1.25 & 1.28  & (9.64 $\pm$ 0.17) $\times$ 10$^{-3}$ \\
\multirow{1}{*}{30--40}   & 0.62 & 0.23 & 0.34 & 1.27 & 0.40 & 0.09 & 0.12 & 0.12 & 0.29 & 1.56 & 1.28  & (6.07 $\pm$ 0.12) $\times$ 10$^{-3}$ \\
\multirow{1}{*}{40--50}   & 0.86 & 0.33 & 0.26 & 0.86 & 0.45 & 0.12 & 0.14 & 0.17 & 0.34 & 1.43 & 1.71  & (3.51 $\pm$ 0.08) $\times$ 10$^{-3}$ \\
\multirow{1}{*}{50--70}   & 1.09 & 0.46 & 0.17 & 1.74 & 0.58 & 0.16 & 0.16 & 0.20 & 0.47 & 2.26 & 1.75  & (1.78 $\pm$ 0.05) $\times$ 10$^{-3}$ \\
\multirow{1}{*}{\x70--110}  & 1.28 & 0.35 & 0.13 & 0.79 & 0.63 & 0.18 & 0.19 & 0.22 & 2.30 & 2.87 & 2.16  & (5.66 $\pm$ 0.20) $\times$ 10$^{-4}$ \\
\multirow{1}{*}{110--150} & 1.44 & 0.51 & 0.14 & 1.37 & 0.62 & 0.20 & 0.22 & 0.25 & 2.31 & 3.18 & 4.46  & (1.45 $\pm$ 0.08) $\times$ 10$^{-4}$ \\
\multirow{1}{*}{150--190} & 1.55 & 1.24 & 0.17 & 1.25 & 0.47 & 0.22 & 0.24 & 0.29 & 4.57 & 5.18 & 7.74  & (4.54 $\pm$ 0.42) $\times$ 10$^{-5}$ \\
\multirow{1}{*}{190--250} & 1.62 & 1.04 & 0.20 & 1.19 & 0.62 & 0.23 & 0.26 & 0.29 & 2.96 & 3.81 & 11.14 & (1.50 $\pm$ 0.18) $\times$ 10$^{-5}$ \\
\multirow{1}{*}{250--600} & 1.65 & 0.62 & 0.20 & 1.78 & 0.66 & 0.23 & 0.27 & 0.34 & 4.07 & 4.85 & 18.07 & (1.18 $\pm$ 0.22) $\times$ 10$^{-6}$ \\
\hline
\end{tabular}
}
\label{tab:ElectronSyst}
\end{center}
\end{table}

\begin{table}[htb]
\renewcommand\arraystretch{1.2}
\topcaption{
The $\PW$ boson normalized differential cross sections for the muon channel in bins of $\ptW$, (1/$\sigma$)(\rd$\sigma$/\rd\pt) ({$\PW \to \mu\PGn$}),
and systematic uncertainties from various sources in units of~\%.
Other details are the same as in Table~\ref{tab:ElectronSyst}.
}
\begin{center}
\resizebox{\textwidth}{!}{
\begin{tabular}{c|c|c|c|c|c|c|c|c|c|c|c|c}
\hline
\multirow{1}{*}{Bin}   &\multirow{1}{*}{Lept.} &\multirow{1}{*}{Mom.}&\multirow{1}{*}{\MET}&\multirow{1}{*}{QCD}  &\multirow{1}{*}{QCD}  &\multirow{2}{*}{EW}&\multirow{1}{*}{SVD} &\multirow{2}{*}{FSR}&\multirow{1}{*}{Unfld.}&\multirow{1}{*}{Total}&\multirow{2}{*}{Stat.}&\multirow{1}{*}{(1/$\sigma$)(d$\sigma$/d\pt)}\\
\multirow{1}{*}{(\GeV)}&\multirow{1}{*}{recon.}&\multirow{1}{*}{res.}&\multirow{1}{*}{res.}&\multirow{1}{*}{bkgr.}&\multirow{1}{*}{shape}&   &\multirow{1}{*}{unfld.}&  &\multirow{1}{*}{bias}&\multirow{1}{*}{syst.}& &\multirow{1}{*}{($\GeV^{-1}$)} \\
\hline
\multirow{1}{*}{\px0--7.5}     & 0.22 & 0.11 & 0.04 & 0.62 & 0.17 & 0.00 & 0.14 & 0.00 & 0.93 & 1.16 & 0.51  & (4.88 $\pm$ 0.06) $\times$ 10$^{-2}$ \\
\multirow{1}{*}{\x7.5--12.5}    & 0.11 & 0.06 & 0.02 & 0.95 & 0.26 & 0.02 & 0.12 & 0.00 & 1.72 & 1.99 & 0.65  & (4.16 $\pm$ 0.09) $\times$ 10$^{-2}$ \\
\multirow{1}{*}{12.5--17.5}   & 0.18 & 0.09 & 0.04 & 0.87 & 0.22 & 0.03 & 0.14 & 0.00 & 1.15 & 1.48 & 0.79  & (2.37 $\pm$ 0.04) $\times$ 10$^{-2}$ \\
\multirow{1}{*}{17.5--24\px}   & 0.32 & 0.20 & 0.06 & 0.94 & 0.27 & 0.04 & 0.17 & 0.00 & 0.30 & 1.11 & 0.85  & (1.43 $\pm$ 0.02) $\times$ 10$^{-2}$ \\
\multirow{1}{*}{24--30}   & 0.40 & 0.25 & 0.06 & 0.94 & 0.28 & 0.02 & 0.18 & 0.00 & 0.65 & 1.28 & 1.14  & (9.25 $\pm$ 0.16) $\times$ 10$^{-3}$ \\
\multirow{1}{*}{30--40}   & 0.38 & 0.24 & 0.06 & 1.52 & 0.26 & 0.03 & 0.19 & 0.01 & 0.27 & 1.64 & 1.14  & (5.91 $\pm$ 0.12) $\times$ 10$^{-3}$ \\
\multirow{1}{*}{40--50}   & 0.31 & 0.17 & 0.06 & 0.89 & 0.15 & 0.06 & 0.21 & 0.01 & 0.44 & 1.09 & 1.58  & (3.50 $\pm$ 0.07) $\times$ 10$^{-3}$ \\
\multirow{1}{*}{50--70}   & 0.29 & 0.14 & 0.07 & 1.47 & 0.31 & 0.10 & 0.26 & 0.01 & 0.78 & 1.74 & 1.57  & (1.77 $\pm$ 0.04) $\times$ 10$^{-3}$ \\
\multirow{1}{*}{\x70--110}  & 0.32 & 0.28 & 0.09 & 0.68 & 0.25 & 0.12 & 0.34 & 0.02 & 1.97 & 2.17 & 2.03  & (5.39 $\pm$ 0.16) $\times$ 10$^{-4}$ \\
\multirow{1}{*}{110--150} & 0.36 & 0.40 & 0.12 & 0.68 & 0.14 & 0.15 & 0.44 & 0.02 & 4.32 & 4.44 & 4.11  & (1.30 $\pm$ 0.08) $\times$ 10$^{-4}$ \\
\multirow{1}{*}{150--190} & 0.39 & 0.49 & 0.15 & 0.70 & 0.62 & 0.16 & 0.53 & 0.02 & 3.07 & 3.32 & 7.89  & (4.21 $\pm$ 0.36) $\times$ 10$^{-5}$ \\
\multirow{1}{*}{190--250} & 0.41 & 0.55 & 0.17 & 0.71 & 0.67 & 0.17 & 0.61 & 0.02 & 5.46 & 5.62 & 12.69 & (1.40 $\pm$ 0.19) $\times$ 10$^{-5}$ \\
\multirow{1}{*}{250--600} & 0.44 & 0.58 & 0.18 & 0.72 & 0.67 & 0.18 & 0.66 & 0.02 & 4.94 & 5.14 & 19.67 & (1.15 $\pm$ 0.23) $\times$ 10$^{-6}$ \\
\hline
\end{tabular}
}
\label{tab:MuonSyst}
\end{center}
\end{table}

\begin{table}[htb]
\renewcommand\arraystretch{1.2}
\topcaption{
The $\Z$ boson normalized differential cross sections for the muon channel in bins of $\ptZ$, (1/$\sigma$)(d$\sigma$/d\pt) ({$\Z \to \PGmp \PGmm$}),
and systematic uncertainties from various sources in units of~\%.
Other details are the same as in Table~\ref{tab:ElectronSyst}.
}
\begin{center}
\begin{tabular}{c|c|c|c|c|c|c|c|c}
\hline
Bin    & \multirow{2}{*}{Bkg.} & Muon & Mom. & Unfld.   & \multirow{2}{*}{FSR} & Total & \multirow{2}{*}{Stat.} & (1/$\sigma$)(d$\sigma$/d\pt) \\
(\GeV) &      & \multicolumn{1}{|c|}{recon.}     & res. &bias   &     & syst.   &    & ($\GeV^{-1}$)    \\
\hline
\px0--2.5     & 0.43                & 0.01                        &  0.02                      & 2.71     & 0.03 	& 2.74            & 5.53        & (3.34 $\pm$ 0.21) $\times$ 10$^{-2}$  \\
2.5--5\px     & 0.42                & 0.00                        &  0.02                      & 1.32     & 0.02 	& 1.38            & 4.59        & (5.53 $\pm$ 0.26) $\times$ 10$^{-2}$  \\
\px5--7.5     & 0.41                & 0.00                        &  0.01                      & 0.28     & 0.01 	& 0.50            & 4.79        & (5.19 $\pm$ 0.25) $\times$ 10$^{-2}$  \\
\x7.5--10\px    & 0.29                & 0.00                        &  0.01                      & 1.30     & 0.01 	& 1.34            & 5.78        & (3.86 $\pm$ 0.23) $\times$ 10$^{-2}$  \\
\px10--12.5   & 0.29                & 0.00                        &  0.01                      & 1.43     & 0.01 	& 1.46            & 5.91        & (3.55 $\pm$ 0.22) $\times$ 10$^{-2}$  \\
12.5--15\px   & 0.23                & 0.00                        &  0.00                      & 2.31     & 0.03 	& 2.33            & 7.52        & (2.41 $\pm$ 0.19) $\times$ 10$^{-2}$  \\
\px15--17.5   & 0.15                & 0.00                        &  0.02                      & 1.29     & 0.02 	& 1.30            & 7.59        & (2.25 $\pm$ 0.17) $\times$ 10$^{-2}$  \\
17.5--20\px   & 0.22                & 0.00                        &  0.01                      & 1.63     & 0.04 	& 1.65            & 8.88        & (1.72 $\pm$ 0.15) $\times$ 10$^{-2}$  \\
20--30   & 0.01                & 0.00                        &  0.01                      & 0.41     & 0.02 	& 0.41            & 4.08        & (1.17 $\pm$ 0.05) $\times$ 10$^{-2}$  \\
30--40   & 0.37                & 0.00                        &  0.01                      & 0.56     & 0.00 	& 0.67            & 5.49        & (6.51 $\pm$ 0.36) $\times$ 10$^{-3}$  \\
40--50   & 0.78                & 0.00                        &  0.01                      & 1.03     & 0.01 	& 1.29            & 7.09        & (4.02 $\pm$ 0.29) $\times$ 10$^{-3}$  \\
50--70   & 1.54                & 0.00                        &  0.01                      & 0.26     & 0.02 	& 1.56            & 6.51        & (2.16 $\pm$ 0.14) $\times$ 10$^{-3}$  \\
70--90   & 2.70                & 0.00                        &  0.03                      & 0.37     & 0.04 	& 2.72            & 10.43       & (8.89 $\pm$ 0.96) $\times$ 10$^{-4}$  \\
\x90--110  & 3.51                & 0.00                        &  0.05                      & 0.67     & 0.01 	& 3.57            & 15.67       & (4.10 $\pm$ 0.66) $\times$ 10$^{-4}$  \\
110--150 & 3.54                & 0.00                        &  0.05                      & 1.14     & 0.13 	& 3.72            & 16.74       & (1.65 $\pm$ 0.28) $\times$ 10$^{-4}$  \\
150--190 & 2.00                & 0.01                        &  0.01                      & 0.14     & 0.18 	& 2.01            & 24.67       & (7.65 $\pm$ 1.89) $\times$ 10$^{-5}$  \\
190--250 & 6.13                & 0.01                        &  0.14                      & 9.91     & 0.33 	& 11.66           & 68.85       & (8.98 $\pm$ 6.27) $\times$ 10$^{-6}$  \\
250--600 & 2.03                & 0.00                        &  0.04                      & 0.45     & 0.23 	& 2.09            & 44.11       & (4.44 $\pm$ 1.96) $\times$ 10$^{-6}$  \\
\hline
\end{tabular}
\label{tab:ZMNormSystFid}
\end{center}
\end{table}

\begin{figure}[htb]
\begin{center}
\includegraphics[width=0.48\textwidth]{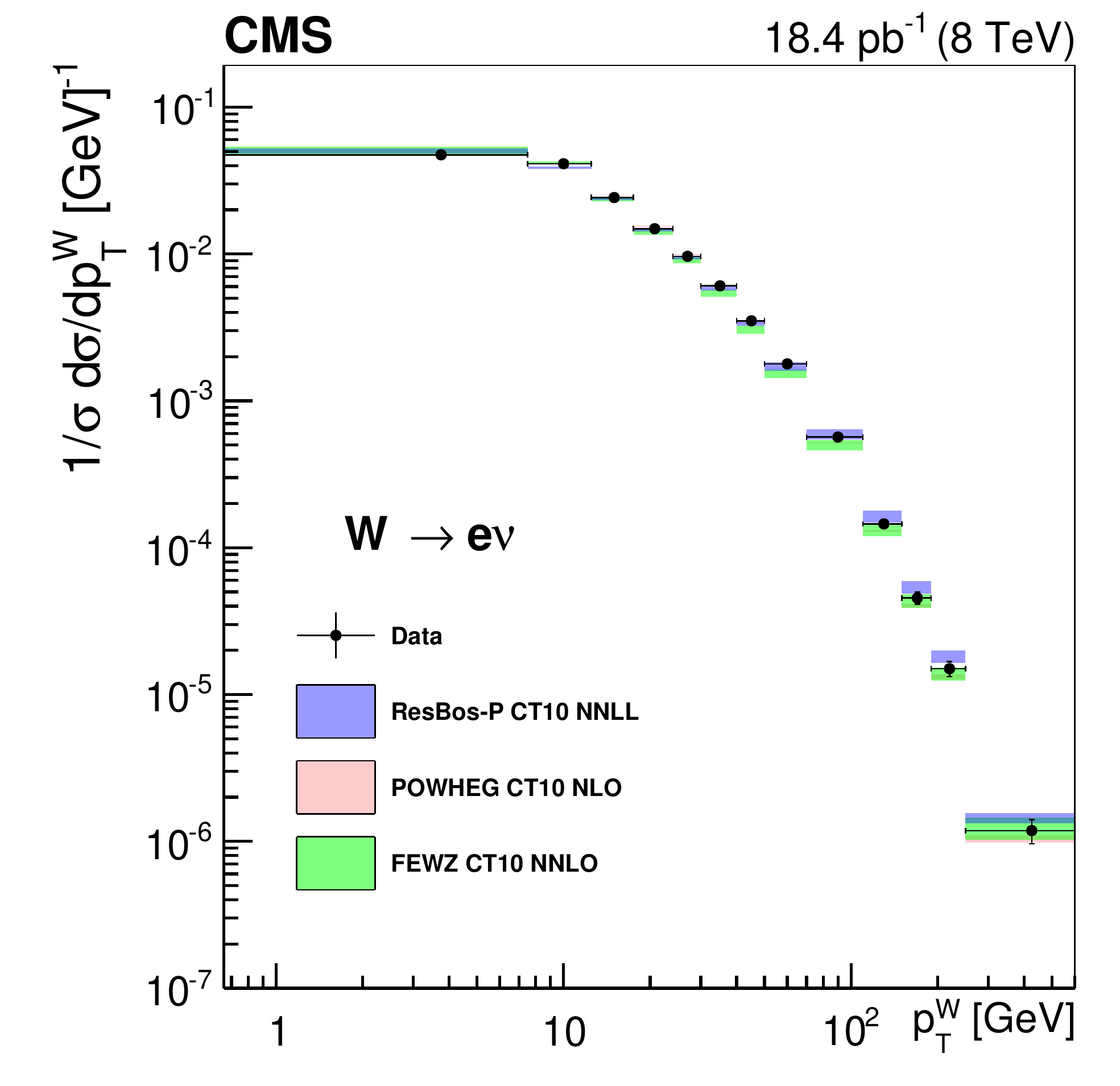}
\includegraphics[width=0.48\textwidth]{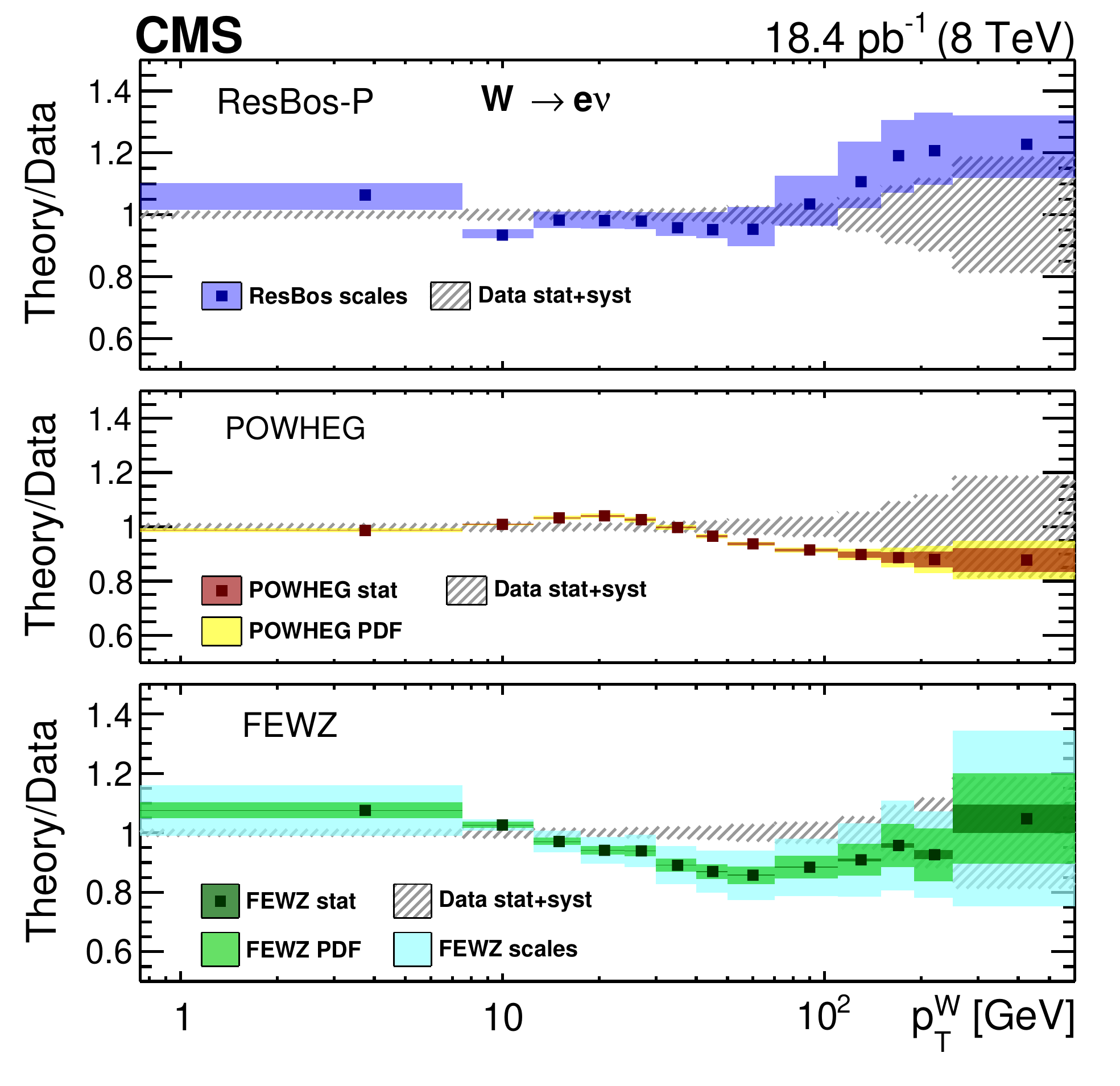}
\includegraphics[width=0.48\textwidth]{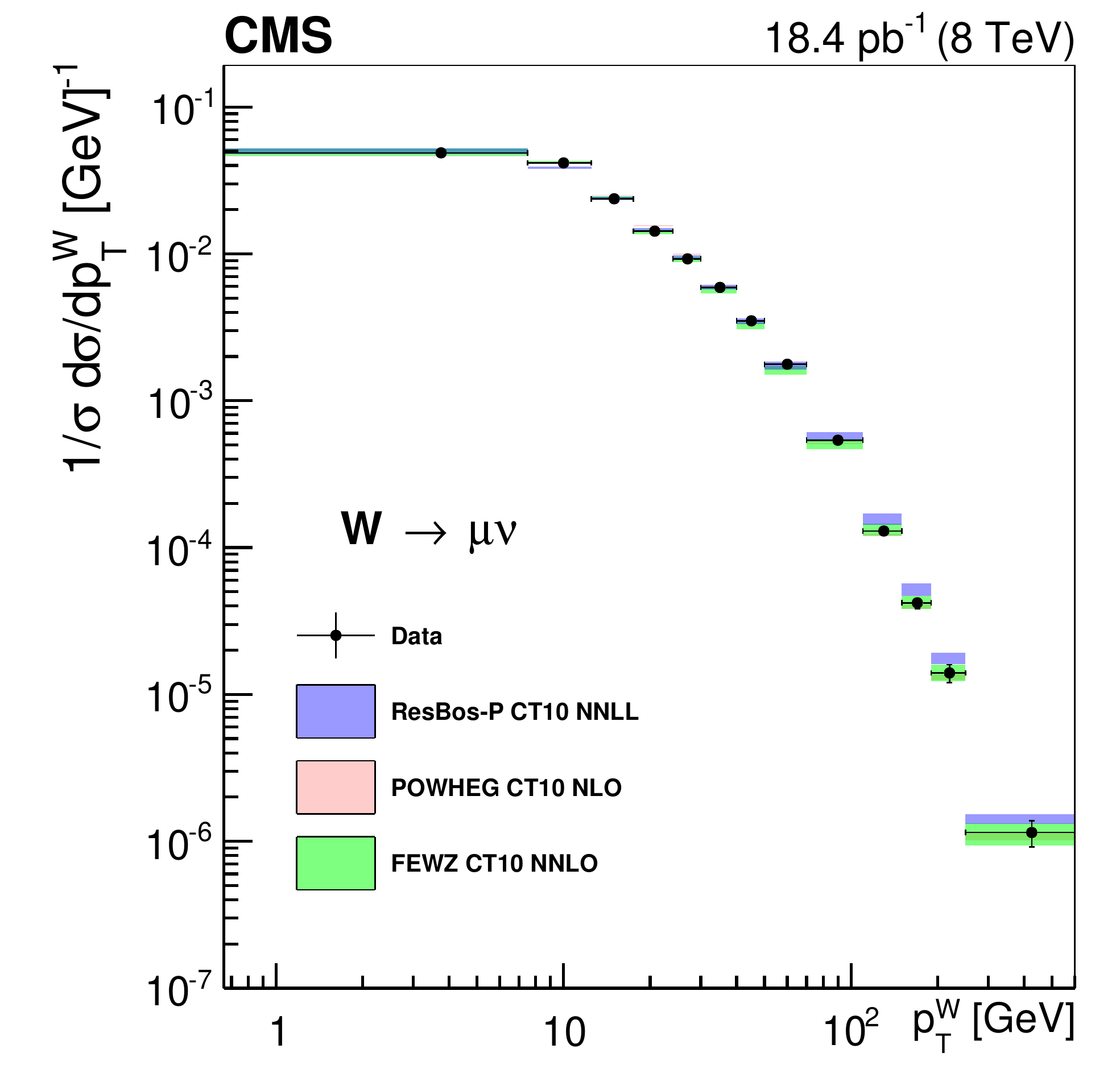}
\includegraphics[width=0.48\textwidth]{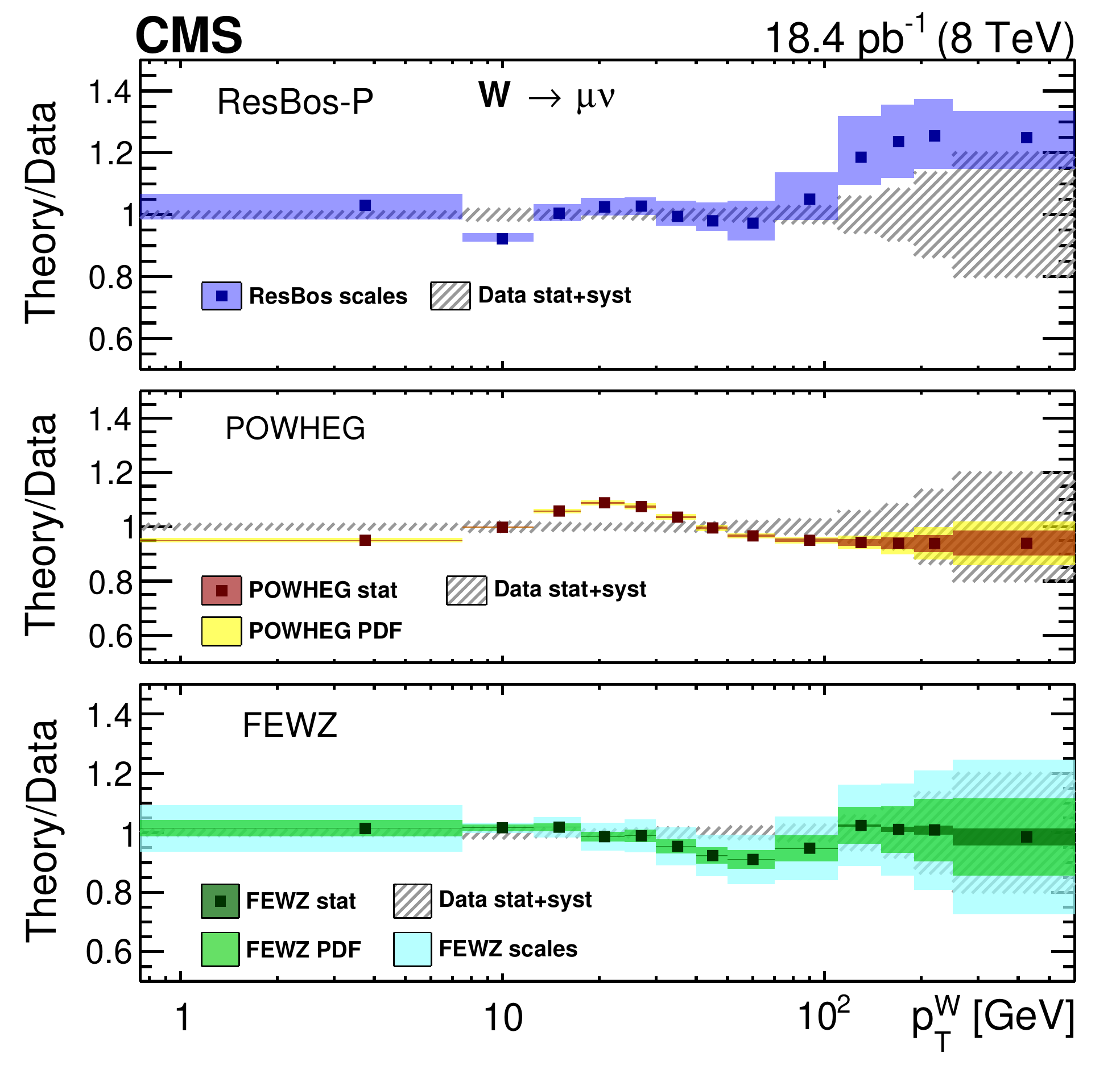}
\caption{\label{fig:Wincl_result}
Normalized differential cross sections for charge independent $\PW$ boson production at the lepton pre-FSR level as a function of $\ptW$ for electron (upper) and muon (lower) decay channels.
The right panels show the ratios of theory predictions to the data.
The bands include (i) the statistical uncertainties, uncertainties from scales, and PDF uncertainties for FEWZ; (ii) the statistical uncertainties and PDF uncertainties for POWHEG;
(iii) the uncertainty from scales for \textsc{ResBos}-P; and (iv) the sum of the statistical and systematic uncertainties in quadrature for data.
}
\end{center}
\end{figure}

The results are compared to three different theoretical predictions: \textsc{ResBos},
\POWHEG , and \FEWZ using CT10~\cite{CT10NLO} PDFs with uncertainties estimated by the method described in Ref.~\cite{LHAPDFUse}.
The resulting spectra for the $\PW$ boson normalized differential cross section are shown in Fig.~\ref{fig:Wincl_result}.

\POWHEG with \PYTHIA using the Z2* tune shows good agreement with the data in the low- and high-$\ptW$ regions,
but overestimates the yield by up to 12\% in the transition region at around 25\GeV.

\textsc{ResBos}-P expectations are consistent with the data for $12.5 < \ptW < 110\GeV$.
Yields are underpredicted for $7.5 < \ptW < 12.5\GeV$.
Above 110\GeV, the predictions systematically overestimate the data by approximately 20\%.

\FEWZ calculates the cross section for gauge boson production at hadron colliders
through order $\mathcal{O}(\alpha^2_s)$ in perturbative QCD.
The $\ptW$ distribution is generated by \FEWZ using perturbative QCD at NNLO.
The CT10 NNLO PDF set is used with dynamic renormalization and factorization scales set to the value of $\sqrt{\smash[b]{M_{\PW}^2 + (\ptW)^2}}$.
The uncertainty of the CT10 PDF set
is numerically propagated through \FEWZ generation.
Scale variations by factors of 1/2 and 2 are applied to estimate the uncertainty. The predictions of \FEWZ are in agreement with the data across the whole range in $\ptW$ within large theoretical uncertainties, except around 60\GeV where it shows 10\% discrepancy.

\begin{figure}[htb]
  \begin{center}
  \includegraphics[width=0.48\textwidth]{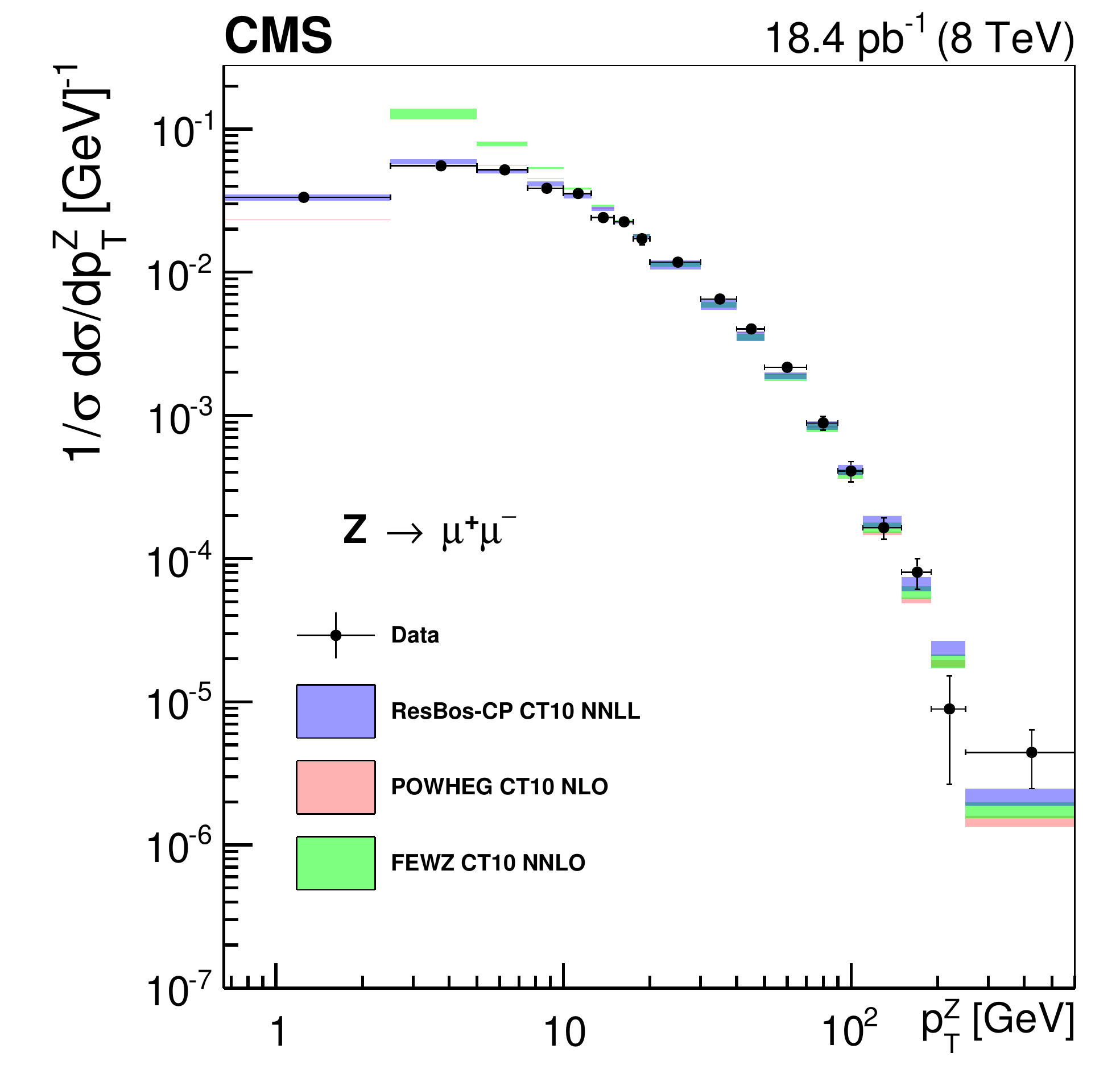}
  \includegraphics[width=0.48\textwidth]{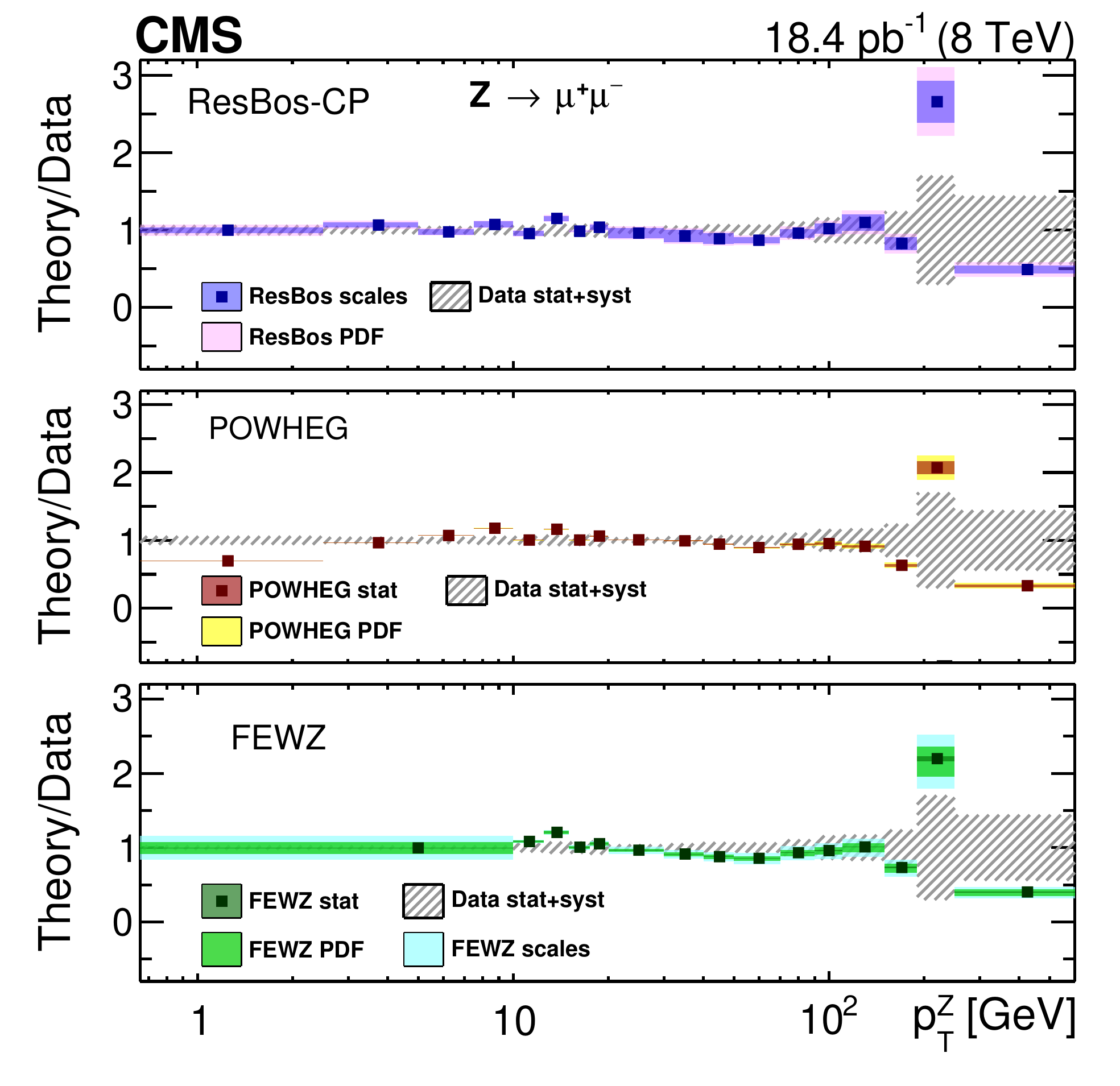}
  \caption{ Comparison of the normalized dimuon differential transverse momentum distribution
    from data (solid symbols) with different theoretical predictions.
    The right panels show the ratios of theory predictions to the data.
    The \textsc{ResBos}-CP version with scale and PDF variation is used for comparison.
  }
  \label{fig:ZpTcombo}
  \end{center}
  \end{figure}

The results for the $\Z$ boson differential cross section are presented in Fig.~\ref{fig:ZpTcombo}.
The \textsc{ResBos}-CP prediction shows good agreement with data in the accessible region of $\ptZ$, whereas \POWHEG shows 30\% lower expectation in the range 0--2.5\GeV and 18\% excess for the interval 7.5--10\GeV.
As anticipated, the \FEWZ prediction with fixed-order perturbation theory shows divergent behavior in the low $\ptZ$ bins ($\ptZ \lesssim 20\GeV$).
A self-consistent test of \FEWZ generation is fulfilled by cross section comparison of the low, high, and full $\ptZ$ region of the measurement.
The ratio of the sum of 0--20 and 20--600\GeV to 0--600\GeV is unity within 10\% uncertainty.
The ratio of the expectation to data at 0--20\GeV is $1.02 \pm 2.6\% (\FEWZ) \pm 1.1\%\,\text{(data)}$.

\subsection{Ratios of the cross sections}

\begin{figure}[htb]
  \begin{center}
  \includegraphics[width=0.48\textwidth]{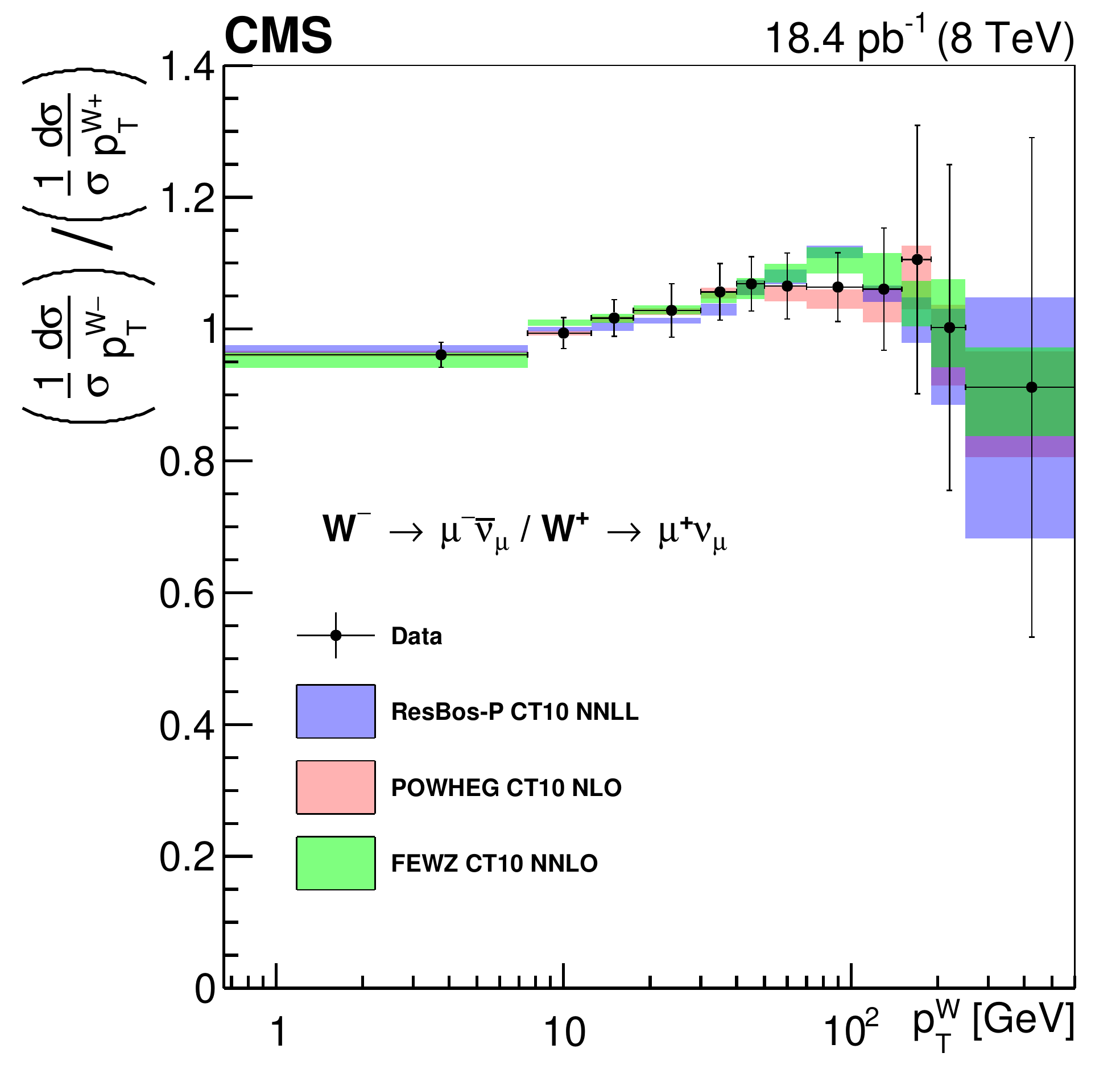}
  \includegraphics[width=0.48\textwidth]{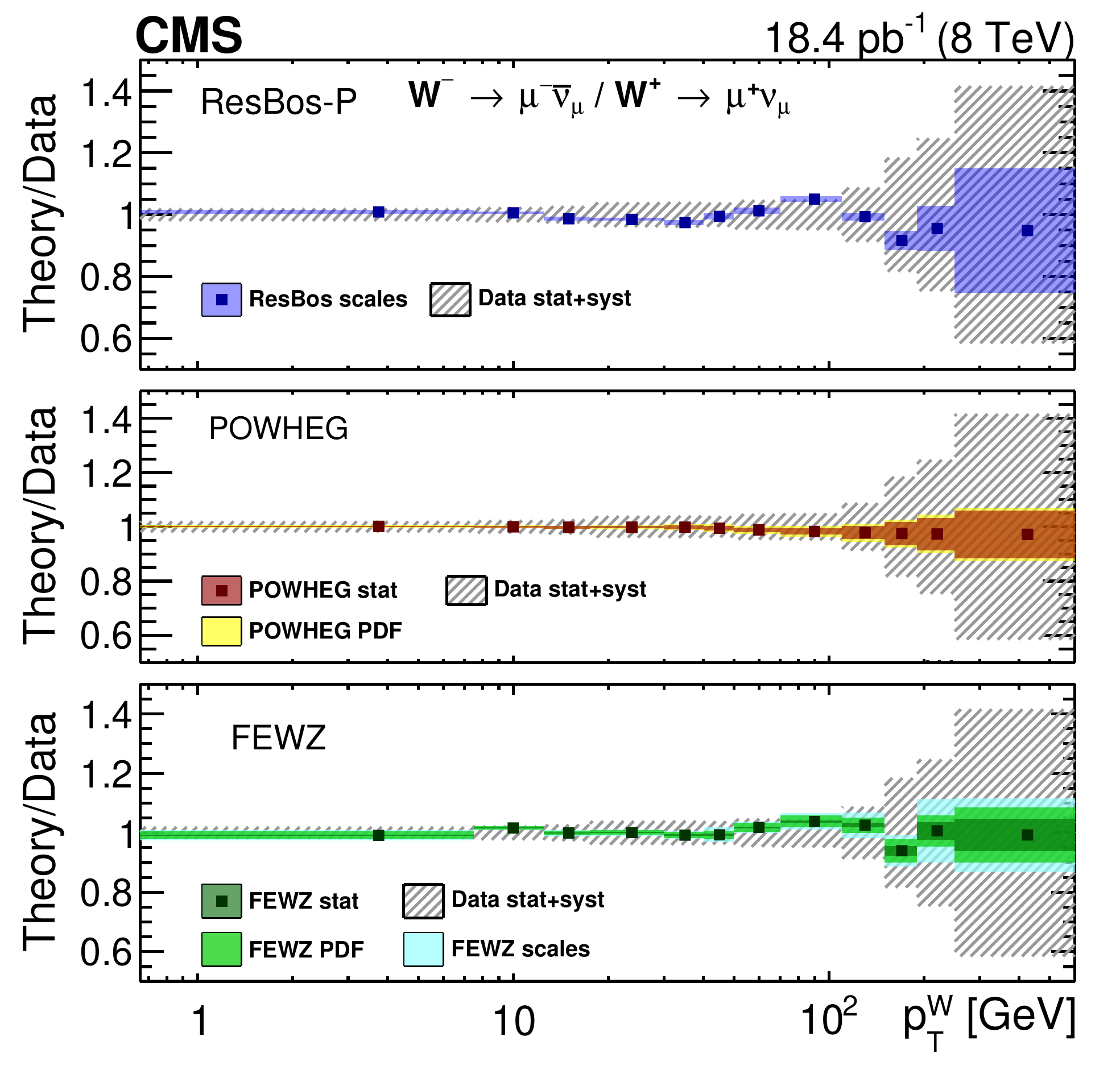}
  \caption{The normalized $\pt$ differential cross section ratio of $\PWm$ to $\PWp$ for muon channel compared with theoretical predictions.
  Data points include the sum of the statistical and systematic uncertainties in quadrature.
  More details are given in the Fig.~\ref{fig:Wincl_result} caption.
  }
  \label{fig:RatiosWmWp}
  \end{center}
  \end{figure}

The ratios of the measured cross sections provide a powerful test of the accuracy of different theoretical
predictions because of full or partial cancellation of theoretical uncertainties.
The ratio of the normalized spectra corresponding to
$\PWm \to \PGmm {\PAGn}$ and $\PWp \to \PGmp \PGn$ decays is shown  in Fig.~\ref{fig:RatiosWmWp}.
The statistical uncertainties in different $\ptV$ bins are considered to be uncorrelated.
The systematic uncertainties are calculated by the method described in Section~\ref{sec:Syst} taking into account all correlations between charge-dependent $\PW$ boson cross sections.
The ratios with the total uncertainty are listed in Table~\ref{tab:WWnWZRatio}.
The results are compared to \POWHEG, \textsc{ResBos}, and \FEWZ predictions. The predictions describe the data reasonably well within experimental uncertainties.

\begin{figure}[htb]
   \begin{center}
   \includegraphics[width=0.48\textwidth]{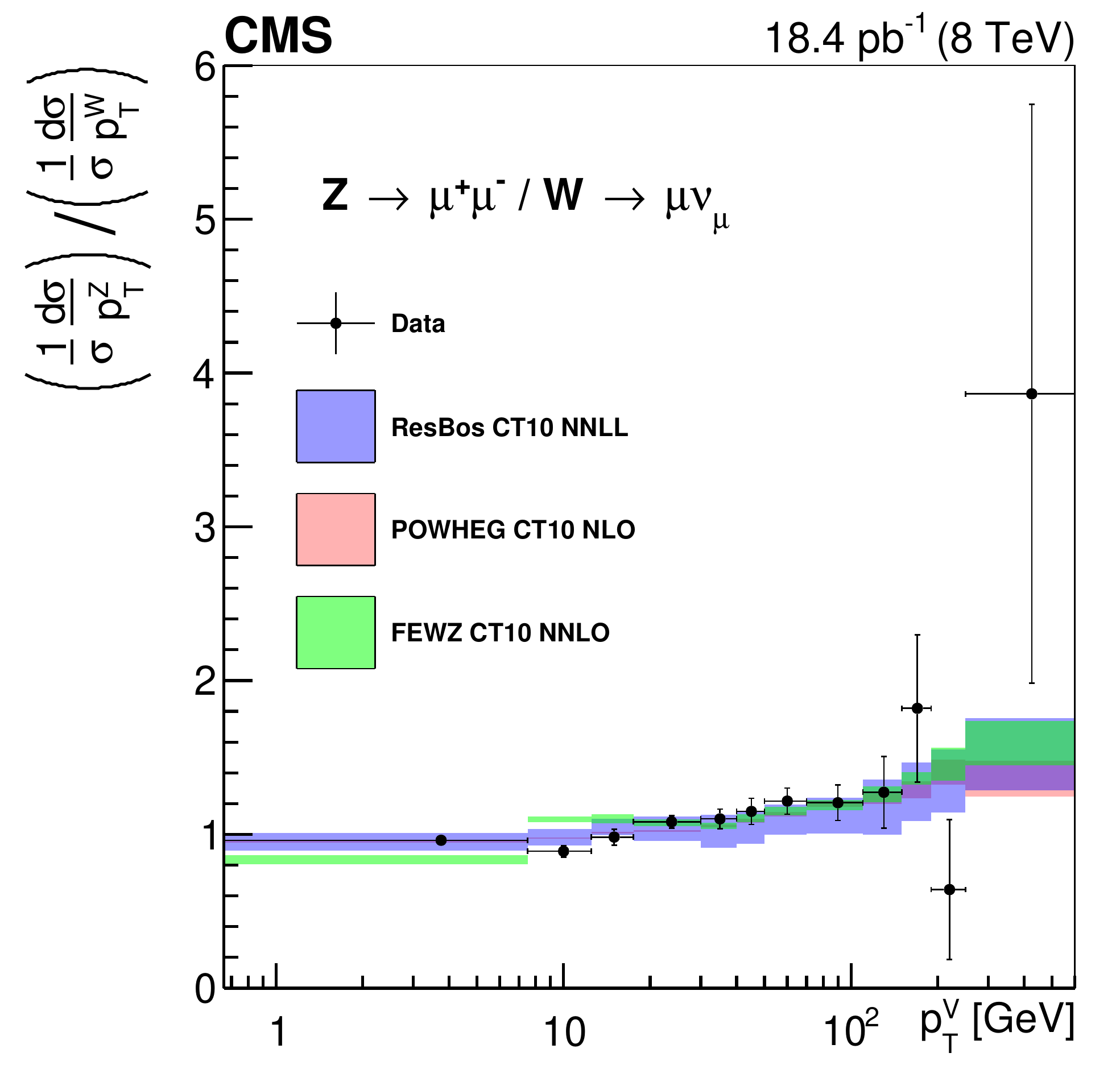}
   \includegraphics[width=0.48\textwidth]{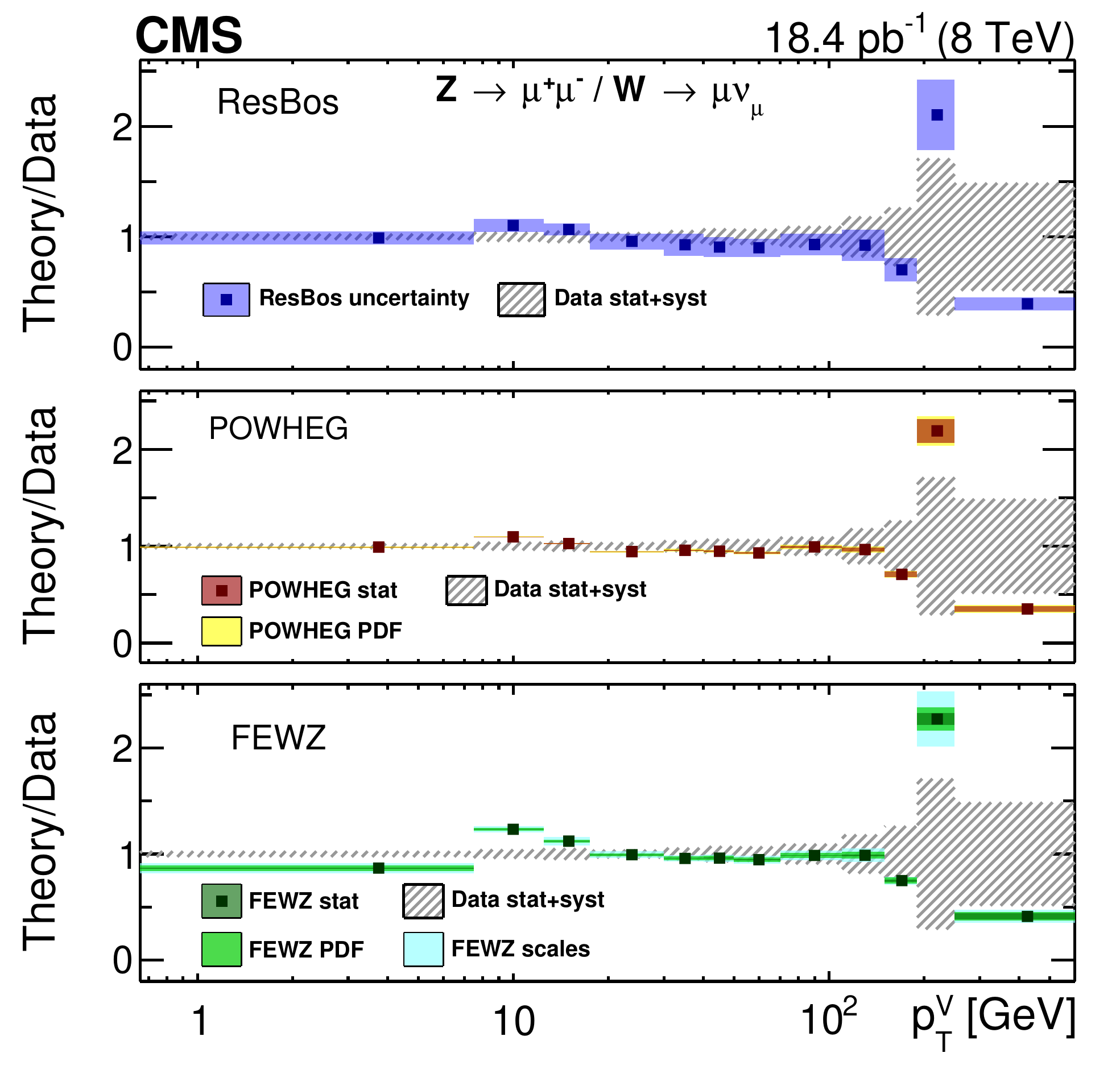}
   \caption{The normalized $\pt$ differential cross section ratio of $\Z$ to $\PW$ for muon channel compared with theoretical predictions.
   The right panels show the ratios of theory predictions to the data.
   The larger than expected uncertainties for \textsc{ResBos} arise from the different strategies in terms of the scale and PDF variations between \textsc{ResBos}-P and \textsc{ResBos}-CP version.
   More details are given in the Fig.~\ref{fig:Wincl_result} and \ref{fig:ZpTcombo} caption.
   }
   \label{fig:RatiosZW}
   \end{center}
   \end{figure}

The ratio of differential production cross sections for $\Z$ to those for $\PW$ in the muon channel is shown in Fig.~\ref{fig:RatiosZW} where the total uncertainties of the measurements are considered to be uncorrelated.
The ratios with the total uncertainty are listed in Table~\ref{tab:WWnWZRatio}.
The \POWHEG calculation shows good agreement with the data in the low- and high-$\ptV$~regions,
 but overestimates the ratio by up to 10\% in the transition region at around $\ptV = 10\GeV$.
The \textsc{ResBos} expectation also shows behavior similar to \POWHEG, but it has larger than expected uncertainties because it employs different strategies in terms of the scale and PDF variations for the $\PW$ and $\Z$ boson generation,
which technically results in no cancellation for their ratio.
\FEWZ predictions describe the data well for $\ptV > 20\GeV$.

\begin{table}[htb]
\renewcommand\arraystretch{1.2}
\topcaption{
Estimated ratios of pre-FSR level normalized differential cross sections within the muon fiducial volume. The uncertainty is the sum of statistical and systematic uncertainties in quadrature.
}
\begin{center}
\begin{tabular}{c|c|c}
\hline
\multirow{1}{*}{Bin (\GeV)}  & \multirow{1}{*}{$\PWm$/$\PWp$} & \multirow{1}{*}{$\Z$/$\PW$} \\
\hline
\multirow{1}{*}{\px0--7.5}     & 0.961 $\pm$ 0.019 & 0.962 $\pm$ 0.025 \\
\multirow{1}{*}{\x7.5--12.5}    & 0.994 $\pm$ 0.024 & 0.890 $\pm$ 0.038 \\
\multirow{1}{*}{12.5--17.5}   & 1.017 $\pm$ 0.028 & 0.982 $\pm$ 0.052 \\
\multirow{1}{*}{17.5--30\px}   & 1.028 $\pm$ 0.041 & 1.081 $\pm$ 0.041 \\
\multirow{1}{*}{30--40}   & 1.056 $\pm$ 0.043 & 1.101 $\pm$ 0.064 \\
\multirow{1}{*}{40--50}   & 1.069 $\pm$ 0.041 & 1.149 $\pm$ 0.085 \\
\multirow{1}{*}{50--70}   & 1.065 $\pm$ 0.050 & 1.216 $\pm$ 0.085 \\
\multirow{1}{*}{\x70--110}  & 1.064 $\pm$ 0.052 & 1.206 $\pm$ 0.115 \\
\multirow{1}{*}{110--150} & 1.061 $\pm$ 0.093 & 1.274 $\pm$ 0.232 \\
\multirow{1}{*}{150--190} & 1.106 $\pm$ 0.204 & 1.820 $\pm$ 0.479 \\
\multirow{1}{*}{190--250} & 1.002 $\pm$ 0.247 & 0.641 $\pm$ 0.454 \\
\multirow{1}{*}{250--600} & 0.912 $\pm$ 0.379 & 3.865 $\pm$ 1.881 \\
\hline

\end{tabular}
\label{tab:WWnWZRatio}
\end{center}
\end{table}

\begin{figure}[htb]
  \begin{center}
    \includegraphics[width=0.48\textwidth]{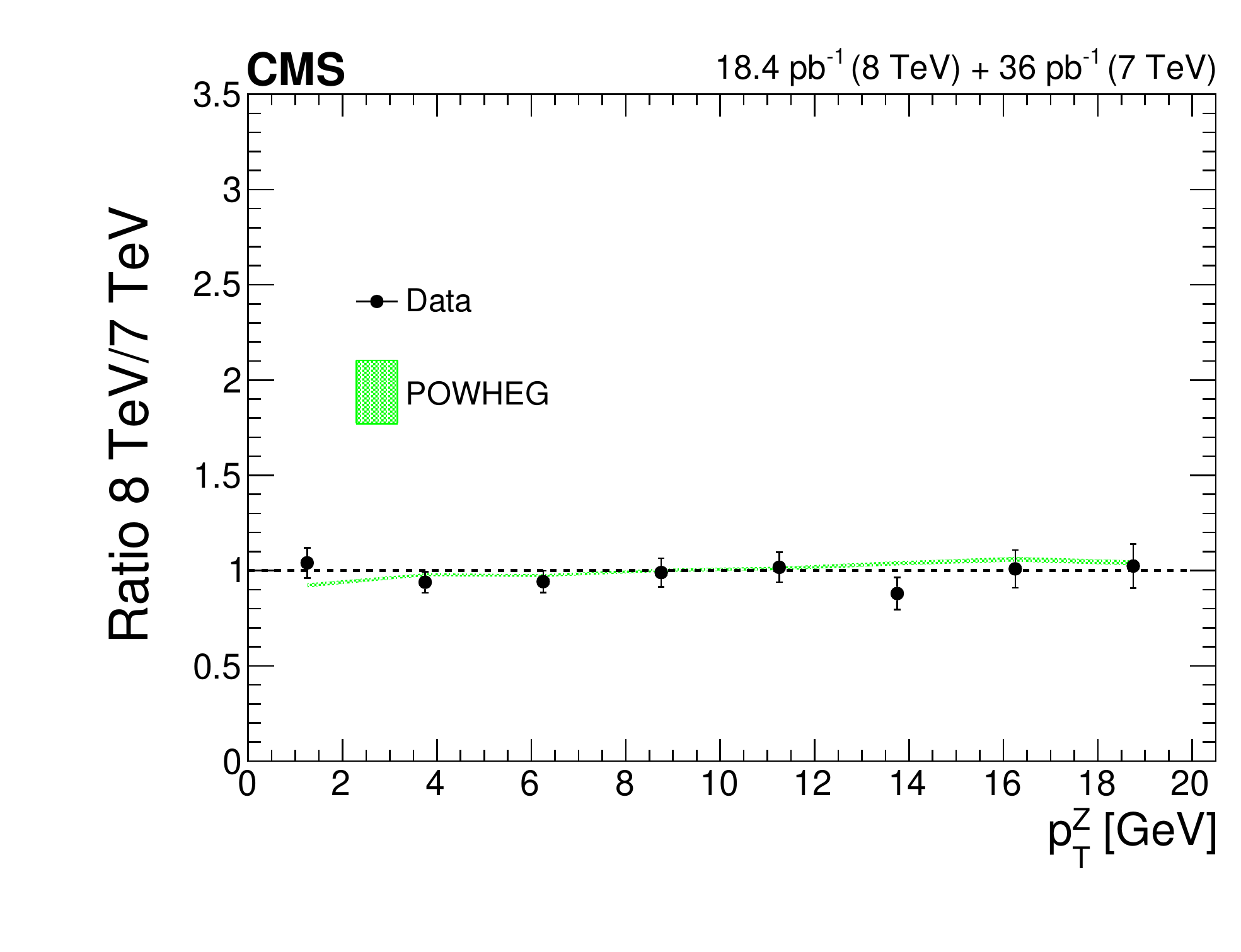}
    \includegraphics[width=0.48\textwidth]{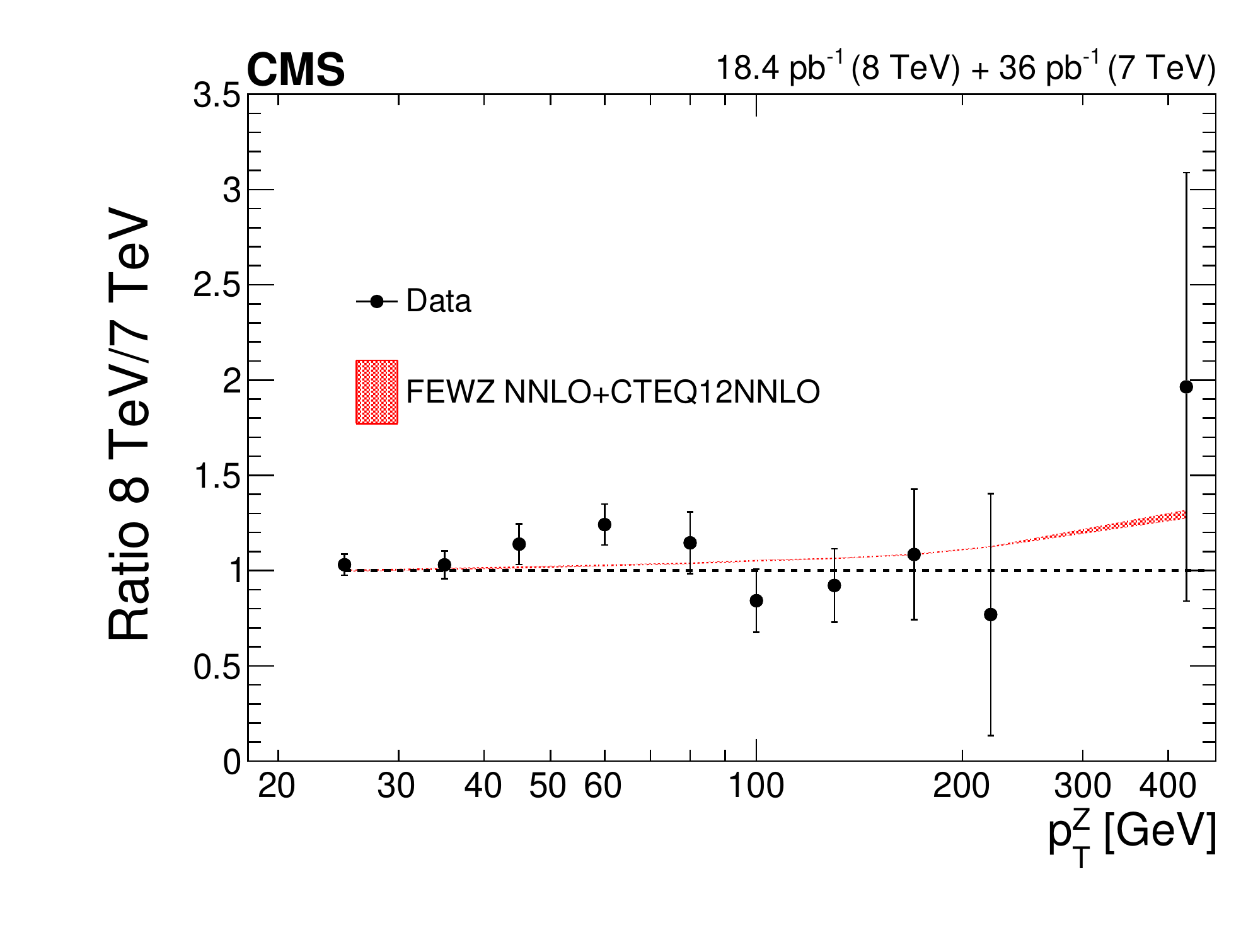}
    \caption{Comparison of the shapes of the differential $\ptZ$ distributions in the muon channel at centre-of-mass energies of 7 and 8\TeV compared with
     the predictions from \POWHEG for $\ptZ < 20\GeV$ and
     \FEWZ for $\ptZ > 20\GeV$.}
    \label{fig:data2012_2010}
  \end{center}
\end{figure}

In Fig.~\ref{fig:data2012_2010} the ratio of differential cross sections for the $\Z$ boson production measured
at two different centre-of-mass energies, 7 and 8\TeV
~\cite{CMS_ZpT7TeV}, are shown for the muon channel, separately for low- and high-$\ptZ$ regions.
The theoretical predictions describe the data well within the experimental uncertainties.

\section{Summary}\label{sec:Summary}

The production cross sections of the weak vector bosons, $\PW$ and $\Z$, as a function of transverse momentum, are measured
by the CMS experiment using a sample of proton-proton collisions during a special low luminosity running of the LHC at $\sqrt{s} = 8\TeV$ that
corresponds to an integrated luminosity of 18.4\pbinv.
The production of $\PW$ bosons is analyzed in both electron and muon decay modes,
while the production of $\Z$ bosons is analyzed using only the dimuon decay channel.

The measured normalized cross sections are compared to various theoretical predictions.
All the predictions provide reasonable descriptions of the data, but \POWHEG at NLO overestimates the yield by up to 12\% around $\ptW$ = 25\GeV.
\POWHEG shows 27\% lower expectation in the $\ptZ$ range 0--2.5\GeV and 18\% excess for the $\ptZ$ interval 7.5--10\GeV.
\FEWZ at NNLO shows 10\% discrepancy around $\ptW$ = 60\GeV and divergent behavior in the low $\ptZ$ region where bin widths are finer than those of the $\PW$ boson study.
\textsc{ResBos}-P systematically overestimates the cross section by approximately 20\% above $\ptW$ = 110\GeV, but the CP version demonstrates good agreement with data in the accessible region of $\ptZ$.
The ratios of $\PWm$ to $\PWp$, $\Z$ to $\PW$ boson differential cross sections, as well as the ratio of $\Z$ boson production cross sections
at centre-of-mass energies 7 to 8\TeV are calculated to allow for more precise comparisons with data.
Overall, the different theoretical models describe the ratios well.

\clearpage
\newpage

\begin{acknowledgments}

\hyphenation{Bundes-ministerium Forschungs-gemeinschaft Forschungs-zentren} We congratulate our colleagues in the CERN accelerator departments for the excellent performance of the LHC and thank the technical and administrative staffs at CERN and at other CMS institutes for their contributions to the success of the CMS effort. In addition, we gratefully acknowledge the computing centres and personnel of the Worldwide LHC Computing Grid for delivering so effectively the computing infrastructure essential to our analyses. Finally, we acknowledge the enduring support for the construction and operation of the LHC and the CMS detector provided by the following funding agencies: the Austrian Federal Ministry of Science, Research and Economy and the Austrian Science Fund; the Belgian Fonds de la Recherche Scientifique, and Fonds voor Wetenschappelijk Onderzoek; the Brazilian Funding Agencies (CNPq, CAPES, FAPERJ, and FAPESP); the Bulgarian Ministry of Education and Science; CERN; the Chinese Academy of Sciences, Ministry of Science and Technology, and National Natural Science Foundation of China; the Colombian Funding Agency (COLCIENCIAS); the Croatian Ministry of Science, Education and Sport, and the Croatian Science Foundation; the Research Promotion Foundation, Cyprus; the Secretariat for Higher Education, Science, Technology and Innovation, Ecuador; the Ministry of Education and Research, Estonian Research Council via IUT23-4 and IUT23-6 and European Regional Development Fund, Estonia; the Academy of Finland, Finnish Ministry of Education and Culture, and Helsinki Institute of Physics; the Institut National de Physique Nucl\'eaire et de Physique des Particules~/~CNRS, and Commissariat \`a l'\'Energie Atomique et aux \'Energies Alternatives~/~CEA, France; the Bundesministerium f\"ur Bildung und Forschung, Deutsche Forschungsgemeinschaft, and Helmholtz-Gemeinschaft Deutscher Forschungszentren, Germany; the General Secretariat for Research and Technology, Greece; the National Scientific Research Foundation, and National Innovation Office, Hungary; the Department of Atomic Energy and the Department of Science and Technology, India; the Institute for Studies in Theoretical Physics and Mathematics, Iran; the Science Foundation, Ireland; the Istituto Nazionale di Fisica Nucleare, Italy; the Ministry of Science, ICT and Future Planning, and National Research Foundation (NRF), Republic of Korea; the Lithuanian Academy of Sciences; the Ministry of Education, and University of Malaya (Malaysia); the Mexican Funding Agencies (BUAP, CINVESTAV, CONACYT, LNS, SEP, and UASLP-FAI); the Ministry of Business, Innovation and Employment, New Zealand; the Pakistan Atomic Energy Commission; the Ministry of Science and Higher Education and the National Science Centre, Poland; the Funda\c{c}\~ao para a Ci\^encia e a Tecnologia, Portugal; JINR, Dubna; the Ministry of Education and Science of the Russian Federation, the Federal Agency of Atomic Energy of the Russian Federation, Russian Academy of Sciences, and the Russian Foundation for Basic Research; the Ministry of Education, Science and Technological Development of Serbia; the Secretar\'{\i}a de Estado de Investigaci\'on, Desarrollo e Innovaci\'on and Programa Consolider-Ingenio 2010, Spain; the Swiss Funding Agencies (ETH Board, ETH Zurich, PSI, SNF, UniZH, Canton Zurich, and SER); the Ministry of Science and Technology, Taipei; the Thailand Center of Excellence in Physics, the Institute for the Promotion of Teaching Science and Technology of Thailand, Special Task Force for Activating Research and the National Science and Technology Development Agency of Thailand; the Scientific and Technical Research Council of Turkey, and Turkish Atomic Energy Authority; the National Academy of Sciences of Ukraine, and State Fund for Fundamental Researches, Ukraine; the Science and Technology Facilities Council, UK; the US Department of Energy, and the US National Science Foundation.

Individuals have received support from the Marie-Curie programme and the European Research Council and EPLANET (European Union); the Leventis Foundation; the A. P. Sloan Foundation; the Alexander von Humboldt Foundation; the Belgian Federal Science Policy Office; the Fonds pour la Formation \`a la Recherche dans l'Industrie et dans l'Agriculture (FRIA-Belgium); the Agentschap voor Innovatie door Wetenschap en Technologie (IWT-Belgium); the Ministry of Education, Youth and Sports (MEYS) of the Czech Republic; the Council of Science and Industrial Research, India; the HOMING PLUS programme of the Foundation for Polish Science, cofinanced from European Union, Regional Development Fund, the Mobility Plus programme of the Ministry of Science and Higher Education, the OPUS programme contract 2014/13/B/ST2/02543 and contract Sonata-bis DEC-2012/07/E/ST2/01406 of the National Science Center (Poland); Kyungpook National University Research Fund (2014) (Republic of Korea); the Thalis and Aristeia programmes cofinanced by EU-ESF and the Greek NSRF; the National Priorities Research Program by Qatar National Research Fund; the Programa Clar\'in-COFUND del Principado de Asturias; the Rachadapisek Sompot Fund for Postdoctoral Fellowship, Chulalongkorn University and the Chulalongkorn Academic into Its 2nd Century Project Advancement Project (Thailand); and the Welch Foundation, contract C-1845.

\end{acknowledgments}
\clearpage
\newpage
\clearpage
\bibliography{auto_generated}

\cleardoublepage \appendix\section{The CMS Collaboration \label{app:collab}}\begin{sloppypar}\hyphenpenalty=5000\widowpenalty=500\clubpenalty=5000\textbf{Yerevan Physics Institute,  Yerevan,  Armenia}\\*[0pt]
V.~Khachatryan, A.M.~Sirunyan, A.~Tumasyan
\vskip\cmsinstskip
\textbf{Institut f\"{u}r Hochenergiephysik der OeAW,  Wien,  Austria}\\*[0pt]
W.~Adam, E.~Asilar, T.~Bergauer, J.~Brandstetter, E.~Brondolin, M.~Dragicevic, J.~Er\"{o}, M.~Flechl, M.~Friedl, R.~Fr\"{u}hwirth\cmsAuthorMark{1}, V.M.~Ghete, C.~Hartl, N.~H\"{o}rmann, J.~Hrubec, M.~Jeitler\cmsAuthorMark{1}, A.~K\"{o}nig, M.~Krammer\cmsAuthorMark{1}, I.~Kr\"{a}tschmer, D.~Liko, T.~Matsushita, I.~Mikulec, D.~Rabady, N.~Rad, B.~Rahbaran, H.~Rohringer, J.~Schieck\cmsAuthorMark{1}, R.~Sch\"{o}fbeck, J.~Strauss, W.~Treberer-Treberspurg, W.~Waltenberger, C.-E.~Wulz\cmsAuthorMark{1}
% ...
\vskip\cmsinstskip
% ...
1:~~Also at Vienna University of Technology, Vienna, Austria\\
% ...

\end{sloppypar}
\end{document}